\documentclass[a4paper, 12pt]{article}

\usepackage{natbib}
\bibpunct{(}{)}{;}{a}{}{,}
\usepackage{setspace}
\usepackage{graphicx}
\usepackage{amsmath, amsthm, amssymb, amsfonts}
\usepackage{url}
\usepackage{multirow}
\usepackage[margin=1in]{geometry}

\makeatletter
\def\url@leostyle{%
  \@ifundefined{selectfont}{\def\UrlFont{\sf}}{\def\UrlFont{\small\ttfamily}}}
\makeatother
\theoremstyle{definition}
\newtheorem{thm}{Theorem}
\theoremstyle{definition}
\newtheorem{lem}{Lemma}
\newtheorem{prop}{Proposition}
\theoremstyle{definition}
\newtheorem{assum}{Assumption}

\theoremstyle{remark}

\theoremstyle{definition}
\newtheorem{defn}{Definition}

\newcommand{\sumi}{\sum_{i=-\infty}^{-1}}

\newcommand{\sumk}{\sum_{k=-\infty}^{\infty}}

\newcommand{\is}{I^*}
\newcommand{\ep}{\epsilon}
\newcommand{\lam}{\lambda}
\newcommand{\sig}{\sigma}
\newcommand{\E}{\mathbb{E}}
\newcommand{\corr}{\mbox{cor}}
\newcommand{\cov}{\mbox{cov}}
\newcommand{\var}{\mbox{var}}
\newcommand{\tr}{\mbox{tr}}
\newcommand{\delt}{\delta_T}
\newcommand{\ept}{\ep_T}
\newcommand{\Delt}{\Delta_T}
\newcommand{\Lamt}{\Lambda_T}

\newcommand{\xtt}{X_{t, T}}
\newcommand{\iitt}{I^{(i)}_{t, T}}
\newcommand{\biz}{\beta_i(z)}

\newcommand{\ystt}{Y^2_{t, T}}
\newcommand{\tystt}{\wt{Y}^2_{t, T}}
\newcommand{\sstt}{\sigma^2_{t, T}}
\newcommand{\zstt}{Z^2_{t, T}}
\newcommand{\ytt}{Y_{t, T}}
\newcommand{\ztt}{Z_{t, T}}
\newcommand{\sst}{\sigma^2(t/T)}

\newcommand{\heta}{\wh{\eta}}
\newcommand{\Y}{\wt{\mathbb{Y}}}
\newcommand{\bS}{\wt{\mathbb{S}}}
\newcommand{\bhS}{\mathbb{S}}
\newcommand{\bbY}{\mathbb{Y}}

\newcommand{\sumsb}{\frac{\sqrt{e-b}}{\sqrt{n}\sqrt{b-s+1}}\sum_{t=s}^b}
\newcommand{\sumbe}{\frac{\sqrt{b-s+1}}{\sqrt{n}\sqrt{e-b}}\sum_{t=b+1}^e}
\newcommand{\sumse}{\sum_{t=s}^e}

\newcommand{\djl}{d_{j, l}}
\newcommand{\hdjl}{\wh{d}_{j, l}}
\newcommand{\mjl}{m_{j, l}}
\newcommand{\sjl}{s_{j, l}}
\newcommand{\ejl}{e_{j, l}}
\newcommand{\njl}{n_{j, l}}
\newcommand{\bjl}{b_{j, l}}
\newcommand{\tjl}{t_{j, l}}

\newcommand{\thr}{\mathit{t}_n}
\newcommand{\td}{\wt{d}}
\newcommand{\tm}{\wt{m}}

\newcommand{\mui}{\mu_{i}}
\newcommand{\muo}{\mu_{1}}
\newcommand{\mut}{\mu_{2}}
\newcommand{\wi}{w_{i}}
\newcommand{\wo}{w_{1}}
\newcommand{\wt}{w_{2}}
\newcommand{\tdi}{\wt{d}_{i}}
\newcommand{\tdo}{\wt{d}_{1}}
\newcommand{\tdt}{\wt{d}_{2}}

\newcommand{\bA}{\mathbf{A}}
\newcommand{\bps}{\mathbf{\psi}}

\def\wh{\widehat}
\def\wt{\widetilde}

\title{
Multiscale and multilevel technique for consistent
segmentation of nonstationary time series}
\author{
Haeran Cho
\thanks{Department of Statistics, London School of Economics, UK.
\protect \ E-mail: {\tt h.cho1@lse.ac.uk}, Phone: +44 (0)20 7955 6014, Fax: +44 (0)20 7955 7416
}
\ and
Piotr Fryzlewicz
\thanks{Department of Statistics, London School of Economics, UK.
\protect \ E-mail: {\tt p.fryzlewicz@lse.ac.uk}
}
}

\date{}

\begin{document}

\maketitle

\begin{center}
\textbf{Abstract}
\end{center}

In this paper, we propose a fast, well-performing, and consistent
method for segmenting a piecewise-stationary, linear time series
with an unknown number of breakpoints. The time series model we use
is the nonparametric Locally Stationary Wavelet model, in which a
complete description of the piecewise-stationary second-order
structure is provided by wavelet periodograms computed at multiple
scales and locations. The initial stage of our method is a new
binary segmentation procedure, with a theoretically justified and
rapidly computable test criterion that detects breakpoints in
wavelet periodograms separately at each scale. This is followed
by  within-scale and across-scales post-processing steps, leading to
consistent estimation of the number and locations of breakpoints in
the second-order structure of the original process.
An extensive simulation study demonstrates good performance of our
method.

\vspace{10pt} \textbf{keywords:} binary segmentation, breakpoint
detection, locally stationary wavelet model, piecewise stationarity,
post-processing, wavelet periodogram.

\section{Introduction}
\label{sec:introduction}

A stationarity assumption is appealing when analysing short time
series. But, it is often unrealistic, for
example when observing time series evolving in naturally
nonstationary environments. One such example can be found in
econometrics, where price processes are considered to have
time-varying variance in response to events taking place in the
market; \citet{mikosch1999}, \citet{leipus2000}, and \citet{starica2005}, among others, 
argued in favour of nonstationary modelling of financial returns. 
For example, given the explosion of market volatility during the 
recent financial crisis, it is unlikely that the same stationary 
time series model can accurately describe the evolution of market
prices before and during the crisis.

Piecewise stationarity is a well studied and arguably the simplest 
form of departure from stationarity, and one
task of interest is to detect breakpoints in the dependence structure.
Breakpoint detection has received considerable attention and the methods that have been developed 
can be broadly categorized into retrospective (a posteriori) methods and on-line methods. 
In the interest of space, we do not review on-line breakpoint 
detection approaches here but refer the reader to \citet{lai2001}.

The ``a posteriori'' approach takes into account
the entire set of observations at once and detects breakpoints which
occurred in the past. Our interest here lies in the ``a posteriori'' segmentation category, 
and we propose a retrospective segmentation procedure that achieves consistency in identifying
multiple breakpoints for a class of nonstationary processes.
(Note that we use the term ``segmentation'' interchangeably with ``multiple breakpoint detection''.)

Early segmentation literature was mostly devoted to testing the
existence of a single breakpoint in the mean or variance of
independent observations
(\citet{chernoff1964}, \citet{sen1975}, \citet{hawkins1977}, \citet{hsu1977}, \citet{worsley1986}).
When the presence of more than one breakpoint is suspected, an algorithm
for detecting multiple breakpoints is needed. 
In \citet{vostrikova1981}, a ``binary segmentation'' procedure was
introduced, a computationally efficient multilevel breakpoint
detection procedure that recursively locates and tests for multiple
breakpoints, producing consistent breakpoint estimators for a class
of random processes with piecewise constant means. However, the
critical values of the tests at each stage are difficult to compute
in practice due to stochasticity in previously selected
breakpoints. \citet{venkatraman1993} employed the same procedure to
find multiple breakpoints in the mean of independent and normally
distributed variables and showed the consistency of the detected
breakpoints with the tests depending on the sample size only, and
thus are easier to compute. The binary segmentation procedure was
also adopted to detect multiple shifts in the variance of
independent observations (\citet{inclan1994}, \citet{chen1997}).

Various multiple breakpoint detection methods have been proposed for
time series of dependent observations. 
In \citet{lavielle2000},
least squares estimators of breakpoint locations were developed for
linear processes with changing mean, extending the work of \citet{bai1998}. 
\citet{adak1998} and \citet{ombao2001} proposed methods that
divided the time series into dyadic blocks and chose the best
segmentation according to suitably tailored criteria.
\citet{whitcher2000,whitcher2002} and \citet{gabbanini2004}
suggested segmenting long memory processes by applying the iterative
cumulative sum of squares (ICSS) algorithm (proposed in
\citet{inclan1994}) to discrete wavelet coefficients of the process
which were approximately Gaussian and decorrelated.
Davis, Lee, and Rodriguez-Yam (2006) developed the Auto-PARM procedure which found the
optimal segmentation of piecewise stationary AR processes via the
minimum description length principle, later extended to the
segmentation of non-linear processes in \citet{davis2008}.
In \citet{lavielle2005}, a breakpoint detection method was developed for weakly or strongly dependent processes with time-varying volatility that minimises a penalised contrast function based on a Gaussian likelihood.
\citet{andreou2002} studied a heuristic segmentation procedure for the GARCH model with changing parameters, 
based on the work of \citet{lavielle2000}.

The aim of our work is to propose a well-performing, theoretically
tractable, and fast procedure for detecting breakpoints in the second-order structure of a piecewise stationary time series
that is linear but otherwise does not follow any
particular parametric model. 
The nonparametric model we use for this purpose is the
Locally Stationary Wavelet (LSW) model first proposed by
Nason, von Sachs and Kroisandt (2000) and later studied by \citet{piotr2006a} and
\citet{bellegem2008}. Detailed justification of our model choice is given in Section \ref{sec:lsw}.
In the LSW model, the piecewise constant second-order structure of the process is completely described by local wavelet periodograms at
multiple scales, and it is those basic statistics that we use as a basis of our segmentation procedure.

To achieve the multiple breakpoint detection, we propose a binary
segmentation method that is applied to wavelet periodograms separately at each scale, 
and followed by a within-scale and across-scales
post-processing procedure to obtain consistent estimators of
breakpoints in the second-order structure of the process. We note
that wavelet periodograms follow a multiplicative statistical model,
but our binary segmentation procedure is different from previously
proposed binary segmentation methods for multiplicative models
(\citet{inclan1994}, \citet{chen1997}) in that it allows for correlated data, which
is essential when working with wavelet periodograms. We note that
Kouamo, Moulines, and Roueff (2010) proposed a CUSUM-type test for detecting a 
{\em single} change in the wavelet variance at one or several 
scales that also permits correlation in the data. We 
emphasise other unique ingredients of our breakpoint detection
procedure which lead to its good performance and consistency in
probability: the theoretical derivation of our test
criterion (which only depends on the length of the time series and
is thus fast to compute); the novel across-scales
post-processing step, essential in combining the results of the
binary segmentation procedures performed for each wavelet
periodogram scale separately.
We note that our method can simultaneously be termed ``multiscale''
and ``multilevel'', as the basic time series model used for our
purpose is wavelet-based and thus is a ``multiscale'' model, and the
core methodology to segment each scale of the wavelet periodogram in
the model is based on binary segmentation and is thus a
``multilevel'' procedure.

The paper is organised as follows. Section \ref{sec:lsw} explains the
LSW model and justifies its choice. Our breakpoint detection
methodology (together with the post-processing steps) is introduced
in Section \ref{sec:second}, where we also demonstrate its theoretical
consistency in estimating the total number and locations of breakpoints. 
In Section \ref{sec:simulation}, we describe the outcome of
an extensive simulation study that demonstrates the good performance of
our method. In Section
\ref{sec:real}, we apply our technique to the segmentation of the Dow
Jones index and this results in the discovery of two
breakpoints: one coinciding with the initial period of the recent
financial crisis, and the other coinciding with the collapse
of Lehman Brothers, a major financial services firm. The proofs of
our theoretical results are provided in the Appendix.
Software (an R script) implementing our methodology is available from:
\url{http://personal.lse.ac.uk/choh1/msml_technique.html}.

\section{Locally stationary wavelet time series}
\label{sec:lsw}

In this section, we define the Locally Stationary Wavelet
(LSW) time series model (noting that our definition is a slight
modification of that of \citet{piotr2006a}), describe its properties,
and justify its choice as an 
attractive framework for developing our time series segmentation methodology.

\begin{defn}
\label{def:lsw} A triangular stochastic array $\{X_{t,
T}\}_{t=0}^{T-1}$ for $T=1, 2, \ldots,$ is in a class of Locally
Stationary Wavelet (LSW) processes if there exists a mean-square
representation
\begin{eqnarray}
\xtt=\sumi\sumk W_i(k/T)\psi_{i, t-k}\xi_{i, k}\label{pswm}
\end{eqnarray}
with $i\in\{-1, -2, \ldots\}$ and $k \in \mathbb{Z}$ as scale and
location parameters, respectively, the $\bps_{i}=(\psi_{i, 0}, \ldots,
\psi_{i, \mathcal{L}_i-1})$ are discrete, real-valued, compactly
supported, non-decimated wavelet vectors with support lengths $\mathcal{L}_i = O(2^{-i})$, 
and the $\xi_{i, k}$ are zero-mean, orthonormal, identically distributed random 
variables. For each $i \le -1$, $W_i(z):[0, 1] \rightarrow \mathcal{R}$ is
a real-valued, piecewise constant function with a finite (but
unknown) number of jumps. If the $L_i$ denote the total magnitude of
jumps in $W_i^2(z)$, the variability of functions $W_i(z)$ is
controlled so that
\begin{itemize}
    \item $\sumi W_i^2(z) < \infty$ uniformly in $z$,
    \item $\sum_{i=-I}^{-1} 2^{-i}L_i = O(\log T)$ where $I=\log_2T$.
\end{itemize}
\end{defn}

The reader unfamiliar with basic concepts in wavelet analysis is
referred to the monograph by \citet{vidakovic1999}.
By way of example, we recall the simplest discrete, non-decimated
wavelet system: the Haar wavelets. Here
\[
\psi_{i,k} = 2^{i/2} \mathbf{I}_{\{0, \ldots, 2^{-j-1}-1\}}(k) - 2^{i/2}\mathbf{I}_{\{2^{-j-1}, \ldots, 2^{-j}-1\}}(k),
\] 
for $i = -1, -2, \ldots$, $k \in \mathbb{Z}$, where $\mathbf{I}_A(k)$ is 1 if $k\in A$ and 0 otherwise.
We note that discrete non-decimated wavelets $\psi_{i, k}$ can be shifted 
to any location defined by the finest-scale wavelets, and not just to `dyadic' locations (i.e. those with shifts being multiples of $2^{-i}$)
as in the discrete wavelet transform. Therefore, discrete 
non-decimated wavelets are no longer an orthogonal, but an overcomplete collection of shifted vectors
(\citet{nason2000}).

Throughout, the $\xi_{i, k}$ are
assumed to follow the normal distribution; extensions to
non-Gaussianity are possible but technically difficult. Comparing the
above definition with the Cram\'{e}r's representation of stationary
processes, $W_i(k/T)$ is a (scale- and location-dependent) transfer
function, the wavelet vectors $\bps_{i}$ are analogous to the
Fourier exponentials, and the innovations $\xi_{i,k}$ correspond to
the orthonormal increment process. Small negative values of the
scale parameter $i$ denote ``fine'' scales where the wavelet vectors
are the most localised and oscillatory; large negative values denote
``coarser'' scales with longer, less oscillatory wavelet vectors. By
assuming that $W_i(z)$ is piecewise constant, we are able to model
processes with a piecewise constant second-order structure where,
between any two breakpoints in $W_i(z)$, the second-order structure
remains constant. 
The Evolutionary Wavelet Spectrum (EWS) is defined as $S_i(z)=W_i(z)^2$, 
and it is in a one-to-one correspondence with the time-dependent 
autocovariance function of the process $c(z, \tau):=\lim_{T\to\infty} \cov(X_{[zT], T}, X_{[zT]+\tau, T})$ (\citet{nason2000}). 
We note that $W_i(z)$ is a valid transfer function; the variance of 
the resulting time series $\xtt$ is uniformly bounded over $t$, 
and the one-to-one correspondence between the autocovariance function 
and $S_i(z)$ leads to model identifiability.
Our objective is to develop a consistent method for detecting breakpoints in the EWS, 
and consequently to provide a segmentation of the original time series. 
The following technical assumption is placed on the breakpoints present in the EWS.

\begin{assum}
\label{assum:one} The set of locations $z$ where (possibly
infinitely many) functions $S_i(z)$ contain a jump, is finite; with
$\mathcal{B}=\{z;\quad \exists\, i \ \lim_{u\rightarrow
z-}S_i(u) \neq \lim_{u\rightarrow z+}S_i(u)\}$, then
$B=|\mathcal{B}|<\infty$.
\end{assum}

\noindent We further define the wavelet periodogram of the LSW time
series.

\begin{defn}
Let $\xtt$ be an LSW process as in (\ref{pswm}). The triangular stochastic array
\begin{eqnarray}
\iitt=\left|\sum_s X_{s, T}\psi_{i, s-t} \right|^2 \label{def:wp}
\end{eqnarray}
is called the wavelet periodogram of $\xtt$ at scale i.
\end{defn}

With the autocorrelation wavelets $\Psi_i(\tau):=\sum_k\psi_{i, k}\psi_{i, k-\tau}$, 
the wavelet operator matrix is defined as $\bA=\left(A_{i, k}\right)_{i, k<0}$ 
with $A_{i, k}:=\langle \Psi_i, \Psi_k\rangle=\sum_\tau\Psi_i(\tau)\Psi_k(\tau)$.
\citet{piotr2006a} showed that the expectation of a wavelet periodogram 
$\E\iitt$ is ``close'' (in the sense of the integrated squared bias converging 
to zero) to the function $\biz=\sum_{j=-\infty}^{-1}S_j(z)A_{i, j}$, a piecewise constant function
with at most $B$ jumps, all of which occur in the set $\mathcal{B}$.
Thus, there exists a one-to-one correspondence between EWS, the time-dependent
autocovariance function, and the function $\biz$ (being the asymptotic expectation of the
wavelet periodogram). Every breakpoint in the autocovariance structure 
then results in a breakpoint in at least one of the $\beta_i(z)$'s, and is thus detectable,
at least with $T\to\infty$, by analysing the wavelet periodogram sequences.
We note that $\E\iitt$ itself is piecewise constant by definition,
except on the intervals of length $C2^{-i}$ around the discontinuities
occurring in $\mathcal{B}$ ($C$ denotes an arbitrary positive constant throughout the paper); 
given a breakpoint $\nu\in\mathcal{B}$, the computation of $\iitt$ for $t\in [\nu-C2^{-i}, \nu+C2^{-i}]$ involves observations from two stationary segments, which results in $\E\iitt$ being ``almost'' piecewise constant yet not completely so.

The finiteness of $\mathcal{B}$ implies that there exists a fixed
index $\is<\lfloor\log_2T\rfloor$ such that each breakpoint in
$\mathcal{B}$ can be found in at least one of the functions $S_i(z)$
for $i=-1, \ldots, -\is$. 
Thus, from the invertibility of $\mathbf{A}$ and the closeness of $\biz$ and $\E\iitt$, as noted above, we conclude that every breakpoint is 
detectable from the wavelet periodogram sequences at scales $i=-1, \ldots, -\is$.
Since $\is$ is fixed but unknown, in our theoretical considerations
we permit it to increase slowly to infinity with $T$, see the Appendix for further details.
A further reason for disregarding the coarse scales $i<-\is$ is that
the autocorrelation within each wavelet periodogram sequence becomes
stronger at coarser scales; similarly, the intervals on which
$\E\iitt$ is not piecewise constant become longer. Thus, for coarse
scales, wavelet periodograms provide little useful information about
breakpoints and can safely be omitted.

We end this section by briefly summarising our reasons behind the
choice of the LSW model as a suitable framework for developing our
methodology:
\begin{itemize}
\item[(i)]
The entire piecewise constant second-order structure of the process is encoded in the (asymptotically) piecewise constant sequences $\E\iitt$.
\item[(ii)]
Due to the ``whitening" property of wavelets, the wavelet periodogram sequences are often much less 
autocorrelated than the original process. In Section 9.2.2 of \citet{vidakovic1999}, the ``whitening'' 
property of wavelets is formalised for a second-order stationary time series $\xtt$ with a sufficiently
smooth spectral density; defining the wavelet coefficient as $r_{i,k}:=\sum_s X_{s, T}\psi_{i, s-k}$, 
the across-scale covariance of the wavelet coefficients $\E(r_{i,k}r_{i',k'})$ vanishes for $|i-i'|>1$,
is arbitrarily small for $|i-i'|=1$, and decays as $o(|k-k'|^{-1})$ within each scale, provided the wavelet 
used is also sufficiently smooth. However, we emphasise that our segmentation method permits autocorrelation 
in the wavelet periodogram sequences, as described later in Section \ref{sec:multiplicative}. 
\item[(iii)]
The entire array of the wavelet periodograms at all scales is easily
and rapidly computable via the non-decimated wavelet transform.
\item[(iv)]
The use of the ``rescaled time'' $z = k/T$ in
(\ref{pswm}) and the associated regularity assumptions on the
transfer functions $W_i(z)$ permit us to establish rigorous
asymptotic properties of our procedure.
\end{itemize}

\section{Binary segmentation algorithm}
\label{sec:second}

Noting that each wavelet periodogram sequence
follows a multiplicative model, as described in Section
\ref{sec:multiplicative}, we introduce a binary segmentation
algorithm for such class of sequences. Binary segmentation is a
computationally efficient tool that searches for multiple
breakpoints in a recursive manner (and can be classed as a
``greedy'' and ``multilevel'' algorithm). \citet{venkatraman1993}
applied the procedure to a sequence of independent normal variables
with multiple breakpoints in its mean and showed that the detected
breakpoints were consistent in terms of their number and locations.
In the following, we aim at extending these consistency results to
the multiplicative model where dependence between
observations is permitted.

\subsection{Generic multiplicative model}
\label{sec:multiplicative}

Recall that each wavelet periodogram ordinate is simply a squared
wavelet coefficient of a zero-mean Gaussian time series, is distributed
as a scaled $\chi^2_1$ variable, and satisfies $\iitt=\E\iitt \cdot
\zstt$, where $\{\ztt\}_{t=0}^{T-1}$ are autocorrelated standard
normal variables. Hence we develop a generic breakpoint
detection tool for multiplicative sequences
\begin{eqnarray}
\ystt=\sstt\cdot\zstt, \ t=0, \ldots, T-1; \label{generic}
\end{eqnarray}

\noindent $\iitt$ and $\E \iitt$ can be viewed as special cases of
$\ystt$ and $\sstt$, respectively. We assume additional conditions that
are, in particular, satisfied for $\iitt$ and $\E \iitt$ by the
assumptions of Theorem 2.

\begin{itemize}
\item[(i)] $\sstt$ is deterministic and ``close'' to a piecewise constant function
$\sigma^2(t/T)$ in the sense that $\sstt$ is piecewise constant
apart from intervals of length at most $C2^{\is}$ around the
discontinuities in $\sigma^2(z)$, and
$T^{-1}\sum_{t=0}^{T-1}|\sstt-\sigma^2(t/T)|^2=o(\log^{-1}T)$, where
the latter rate comes from the rate of convergence of the integrated
squared bias between $\beta_i(t/T)$ and $\E\iitt$ (see
\citet{piotr2006a} for details) and from the fact that our
attention is limited to the $\is$ finest scales only. Further,
$\sigma^2(z)$ is bounded from above and away from zero, with a
finite but unknown number of jumps.
\item[(ii)] $\{\ztt\}_{t=0}^{T-1}$ is a sequence of standard Gaussian variables
and the function $\rho(\tau)=\sup_{t,T}$ $|\corr(Z_{t, T}, Z_{t+\tau, T})|$ satisfies
$\rho^1_{\infty}<\infty$ where
$\rho^p_{\infty}=\sum_{\tau}\left\vert\rho(\tau)\right\vert^p$. 
\end{itemize}
Once the breakpoint detection algorithm for the generic model (\ref{generic})
has been established, we apply it to the wavelet periodograms.

\subsection{Algorithm}
\label{sec:algorithm}

The first step of the binary segmentation procedure is to find the likely location of a breakpoint.
We locate such a point in the interval $(0, T-1)$ as the one which maximizes the absolute value of
\begin{eqnarray}
\bbY_{0, T-1}^{\nu}
=\sqrt{\frac{T-\nu}{T\cdot\nu}}\sum_{t=0}^{\nu-1}\ystt-
\sqrt{\frac{\nu}{T\cdot(T-\nu)}}\sum_{t=\nu}^{T-1}\ystt.
\label{unbalhaar}
\end{eqnarray}
Here $\bbY_{0, T-1}^{\nu}$ can be interpreted as a scaled difference
between the partial means of two segments $\{\ystt\}_{t=0}^{\nu-1}$
and $\{\ystt\}_{t=\nu}^{T-1}$, where the scaling is chosen so as to
keep the variance $\bbY_{0, T-1}^{\nu}$ constant over $\nu$ in the
idealised case of $Y_{t,T}^2$ being i.i.d. Once such a $\nu$ has
been found, we use $\bbY_{0, T-1}^{\nu}$ (but not only this quantity; see below for details) to test the null hypothesis of $\sst$ being constant over $[0, T-1]$. 
The test statistic and its critical value are established such
that when a breakpoint is present, the null hypothesis is rejected
with probability converging to $1$. If the null hypothesis is
rejected, we continue the simultaneous locating and testing of
breakpoints on the two segments to the left and right of
$\nu$ in a recursive manner until no further breakpoints are
detected. The algorithm is summarised below, where $j$ is the level
index and $l$ is the location index of the node at each level.
Here the term ``level'' is used to indicate the progression 
of the segmentation procedure.

\vspace{10pt}

\noindent \textbf{Algorithm}

\begin{description}
\item[Step 1]
Begin with $(j, l)=(1, 1)$. Let $\sjl=0$ and $\ejl=T-1$.

\item[Step 2]
Iteratively compute $\bbY^b_{\sjl, \ejl}$ as in (\ref{unbalhaar})
for $b\in(\sjl, \ejl)$.
Then, find $\bjl$ which maximizes its absolute
value while satisfying 
\[
\max\left\{\sqrt{(\ejl-\bjl)/(\bjl-\sjl+1)}, \sqrt{(\bjl-\sjl+1)/(\ejl-\bjl)}\right\} \le c
\] for a fixed constant $c\in(0, \infty)$. Let $\njl=\ejl-\sjl+1$, $\djl=\bbY^{\bjl}_{\sjl, \ejl}$, and
$\mjl=\sum_{t=\sjl}^{\ejl} \ystt/\sqrt{\njl}$.

\item[Step 3]
Perform hard thresholding on $|\djl|/\mjl$ with the threshold
$\tjl=\tau T^{\theta}\sqrt{\log T/\njl}$ so that
$\hdjl=\djl$ if $|\djl|>\mjl \cdot \tjl$, and $\hdjl=0$ otherwise.
The choice of $\theta$ and $\tau$ is discussed in Section
\ref{sec:parameters}.

\item[Step 4] If either $\hdjl=0$ or $\max\{\bjl-\sjl+1, \ejl-\bjl\}<\Delt$
for $l$, stop the algorithm on the interval $[\sjl, \ejl]$;
if not, let $(s_{j+1, 2l-1}, e_{j+1, 2l-1})=(\sjl, \bjl)$ and
$(s_{j+1, 2l}, e_{j+1, 2l})=(\bjl+1, \ejl)$, and update the level $j$ as $j \rightarrow j+1$.
The choice of $\Delt$ is discussed in Section \ref{sec:parameters}.

\item[Step 5] Repeat Steps 2--4.

\end{description}

The condition imposed on $\bjl$ in Step 2 implies that the breakpoints should 
be sufficiently scattered over time without being too close to each other, and
a similar condition is required of the true breakpoints in $\sst$, see 
Assumption \ref{assum:two} in Section \ref{sec:consistency}.
The set of detected breakpoints is $\{\bjl; \hdjl\ne 0\}$. The test
statistic $|\djl|/\mjl$ is a scaled version of the test statistics
in the ICSS algorithm (\citet{inclan1994}). However, the
test criteria in that paper are derived empirically under the
assumption of independent observations, and there is no guarantee
that their algorithm produces consistent breakpoint estimates. 

\citet{piotr2006a} and \citet{fss06} introduced ``Haar-Fisz'' techniques in 
different contexts; the former for estimating the time-varying local variance of an 
LSW time series, and the latter for estimating time-varying volatility in a 
locally stationary model for financial log-returns. Each Haar-Fisz method has a 
device (termed the ``Fisz transform'') for stabilising the variance of the Haar wavelet
coefficients of the data and thereby bringing the distribution of the data close to 
Gaussianity with constant variance. This is similar to the step in our algorithm where 
the differential statistic ($d_{j, l}$) is divided by the local mean ($m_{j, l}$), with the convention $0/0 = 0$. 
However, the Fisz transform was only defined for the case $b = \frac{1}{2}(e_{j,l} + s_{j,l} + 1)$ (meaning the segments were split
in half) and it was not used for the purposes of breakpoint detection.

\subsubsection{Post-processing within a sequence}
\label{sec:within}

We equip the procedure with an extra step aimed at reducing
the risk of overestimating the number of breakpoints. The ICSS
algorithm in \citet{inclan1994} has a ``fine-tune'' step whereby if
more than one breakpoint is found, each breakpoint is checked
against the adjacent ones to reduce the risk of overestimation. We
propose a post-processing procedure performing a similar task within
the single-sequence multiplicative model (\ref{generic}). At each
breakpoint, the test statistic is re-calculated over the interval
between two neighbouring breakpoints and compared with the threshold. 
Denote the breakpoint estimates as $\heta_p, p=1, \ldots,
\wh{N}$ and $\heta_0=0$,  $\heta_{\wh{N}+1}=T$. For each
$\heta_p$, we examine whether
$
\left| \bbY^{\heta_{p}}_{\heta_{p-1}+1, \heta_{p+1}} \right|>\tau
T^{\theta}\sqrt{\log T} \cdot
\sum_{t=\heta_{p-1}+1}^{\heta_{p+1}}\ystt/(\heta_{p+1}-\heta_{p-1}).
$
If this inequality does not hold, $\heta_{p}$ is removed and
the same procedure is repeated with the reduced set of breakpoints
until the set does not change. 

We emphasise that our within-scale post-processing step is in line with the theoretical derivation of breakpoint detection consistency as 
(a) the extra checks are of the same form as those done in the original algorithm,
(b) the locations of the breakpoints that survive the post-processing are unchanged.
The next section provides details of our consistency result.

\subsection{Consistency of detected breakpoints}
\label{sec:consistency}

In this section, we first show
the consistency of our algorithm for a multiplicative sequence as in
(\ref{generic}), which corresponds to the wavelet periodogram
sequence at a single scale. Later, Theorem \ref{thm:two} shows
how this consistency result carries over to the consistency of our
procedure in detecting breakpoints in the entire second-order
structure of the input LSW process $X_{t,T}$.

Denote the number of breakpoints in $\sst$ by $N$ and the
breakpoints themselves by $0 <\eta_1< \ldots < \eta_N < T-1$, with
$\eta_0=0, \ \eta_{N+1}=T-1$. 

\begin{assum}
\label{assum:two} For $\Theta \in (7/8, 1]$ and $\theta\in(5/4-\Theta, \Theta-1/2)$, 
the length of each segment in $\sigma^2(t/T)$ is bounded from below by
$\delta_T=CT^{\Theta}$. Further, there exists some constant $c\in(0, \infty)$ such that,
\begin{eqnarray}
\max_{1\le p\le N}\left\{\sqrt{\frac{\eta_p-\eta_{p-1}}{\eta_{p+1}-\eta_{p}}},
\sqrt{\frac{\eta_{p+1}-\eta_{p}}{\eta_{p}-\eta_{p-1}}}\right\} \le c.
\nonumber
\end{eqnarray}
\end{assum}

\begin{thm}
\label{thm:one} 
Suppose that $\{\ytt\}_{t=0}^{T-1}$ follows model
(\ref{generic}). Assume there exist $M, m>0$ such that $\sup_t|\sst| \le M$ and $\inf_{1\le i\le N}\left\vert\sigma^2\left((\eta_{i}+1)/T\right)-\sigma^2\left(\eta_{i}/T\right)\right\vert\ge m$. 
Under Assumption \ref{assum:two}, the number and locations of the detected breakpoints are consistent. That is, 
$\mathbf{P}\left\{ \wh{N}=N; \, \left|\heta_p-\eta_p\right| \le C\ept, \ 1\le p \le N \right\}\to1$ as $T \to \infty$, 
where $\heta_p, \ p=1, \ldots, \wh{N}$ are detected breakpoints and
$\ept=T^{5/2-2\Theta}\log T$.  
(Interpreting this in the rescaled time interval $[0,1]$, $\ept/T = T^{3/2-2\Theta}\log T \to 0$ as $T \to 0$.)
\end{thm}

\subsubsection{Post-processing across the scales}
\label{sec:across}

We only consider wavelet
periodograms $\iitt$ at scales $i=-1, \ldots, -\is$, choosing $\is$
to satisfy $2^{\is} \ll \ept=T^{5/2-2\Theta}\log T$, so that the bias between
$\sstt$ and $\sst$ does not preclude the results of
Theorem \ref{thm:one}. Recall that any
breakpoint in the second-order structure of the original process
$X_{t,T}$ must be reflected in a breakpoint in at least one of the
asymptotic wavelet periodogram expectations $\biz, \ i = -1,
\ldots, -I^*$, and vice versa: a breakpoint in one of the $\biz$'s
implies a breakpoint in the second-order structure of $X_{t,T}$.
Thus, it is sensible to combine the estimated breakpoints across the
periodogram scales by, roughly speaking, selecting a breakpoint as
significant if it appears in {\em any} of the wavelet periodogram
sequences. This section provides a precise algorithm for doing
this, and states a consistency result for the final set of
breakpoints arising from combining them across scales.

The complete across-scales post-processing algorithm follows. Denote the set of detected breakpoints from the sequence $\iitt$ as
$\wh{\mathcal{B}}_i=\left\{\heta_p^{(i)}, \ p=1, \ldots,
\wh{N}_i\right\}$. Then the post-processing finds a subset of
$\cup_{i=-1}^{-\is}\wh{\mathcal{B}}_i$,  say $\wh{\mathcal{B}}$, as follows.
\begin{description}
\item[Step 1]
Arrange all breakpoints into groups so that those from different
sequences and within the distance of $\Lamt$ from each other are
classified to the same group; denote the groups by
$\mathcal{G}_1, \ldots, \mathcal{G}_{\wh{B}}$.
\item[Step 2]
Find $i_0=\max\left\{ \arg\max_{-\is \le k \le -1} \wh{N}_k
\right\}$, the finest scale with the most breakpoints.
\item[Step 3]
Check whether there exists $\heta^{(i_0)}_{p_0}$ for every
$\heta^{(i)}_{p}$, $i \ne i_0, \ 1\le p \le \wh{N}_i$, which
satisfies $\left\vert \heta^{(i)}_{p}-\heta^{(i_0)}_{p_0}
\right\vert < \Lamt$. If so, let
$\wh{\mathcal{B}}=\wh{\mathcal{B}}_{i_0}$ and stop the
post-processing.
\item[Step 4]
Otherwise let $\wh{\mathcal{B}}=\left\{\wh{\nu}_p, \ p=1, \ldots,
\wh{B} \right\}$ where each $\wh{\nu}_p\in\mathcal{G}_p$ with the
maximum $i$.
\end{description}
We set $\Lamt=\lfloor\ept/2\rfloor$ in order to take into account
the bias arising in deriving the results of Theorem \ref{thm:one}.
Breakpoints detected at coarser scales are
likely to be less accurate than those detected at finer scales;
therefore, our algorithm prefers the latter. The across-scales
post-processing procedure preserves the number of ``distinct''
breakpoints and also their locations as determined by the algorithm.
Hence the breakpoints in set $\wh{\mathcal{B}}$ are still
consistent estimates of true breakpoints in the second-order
structure of the original nonstationary process $X_{t,T}$.
Although this is not the only way to combine the breakpoints across scales consistent with our theory, we advocate it
due to its good performance.

Denote the set of the true breakpoints in the second-order structure
of $X_{t,T}$ by $\mathcal{B}=\left\{\nu_p, \ p=1, \ldots, B
\right\}$, and the estimated breakpoints by
$\wh{\mathcal{B}}=\left\{\wh{\nu}_p, \ p=1, \ldots, \wh{B}
\right\}$.

\begin{thm}
\label{thm:two}
Suppose that $X_{t,T}$ satisfies Assumption \ref{assum:one} and
that $\nu_p, 1\le p \le B$ satisfy the condition required
of the $\eta_p$'s in Assumption \ref{assum:two}.
Further assume that the conditions on $\sigma^2(z)$ in Theorem \ref{thm:one} hold for each $\biz$.
Then 
$
\mathbf{P}\left\{\wh{B}=B; \ \left|\wh{\nu}_p-\nu_p\right|\le C\ept, \
1\le p \le B \right\} \to 1 
$
as $T \to \infty$.
\end{thm}

\subsection{Choice of $\Delt$, $\theta$, $\tau$ and $\is$}
\label{sec:parameters}

To ensure that each estimated segment is of sufficiently large
length so as not to distort our theoretical results, $\Delt$ is
chosen so that $\Delt \ge C\ept$. In practice our method
works well for smaller values of $\Delt$ as well, and in the
simulation experiments, $\Delt=C\sqrt{T}$ is used. As
$\theta \in (1/4, 1/2)$, we use $\theta=0.251$ (we have found
that the method works best when $\theta$ is close to the lower end
of its permitted range) and elaborate on the choice of $\tau$ below.

The selection of $\tau$ is not a straightforward task and to get
some insight into the issue, a set of numerical experiments was
conducted. A vector of random variables
$\mathbf{X}\sim\mathcal{N}_T(0, \Sigma)$ was generated,
$\mathbf{X}=(X_1, \ldots, X_T)^T$, then transformed into
sequences of wavelet periodograms $\iitt$. The covariance matrix
satisfied $\Sigma=\left(\sig_{i, j}\right)_{i, j=1}^T$ where
$\sig_{i, j}=\rho^{|i-j|}$.
Then we found $b \in (1, T)$ that maximised
\[
\mathbb{I}^b_i=\left\vert\sqrt{\frac{T-b}{T\cdot b}}\sum_{t=1}^b
\iitt-\sqrt{\frac{b}{T(T-b)}}\sum_{t=b+1}^T \iitt \right\vert,
\] and computed
$\mathbb{U}_{i, \rho, T}=\mathbb{I}^b_i\cdot \{T^{-1}\sum_{t=1}^T
\iitt\cdot T^{\theta}\sqrt{\log T}\}^{-1}$. This was repeated with a
varying covariance matrix ($\rho=0, 0.3, 0.6, 0.9$) and sample size
($T=512, 1024, 2048$), 100 times for each combination.

The quantity $\mathbb{U}_{i, \rho, T}$ is the ratio between our test
statistic and the time-dependent factor
$T^{\theta}\sqrt{\log T}$ appearing in the threshold defined in the
algorithm of Section \ref{sec:algorithm}. $\mathbb{U}_{i, \rho, T}$
is computed under the ``null hypothesis'' of no breakpoints being
present in the covariance structure of $X_{t}$, and its magnitude
serves as a guideline as to how to select the value of $\tau$, for
each scale $i$, to prevent spurious breakpoint detection in the null
hypothesis case. The results showed that the values of
$\mathbb{U}_{i, \rho, T}$ and their range tended to increase for
coarser scales, this due to the increasing dependence in the wavelet
periodogram sequences. In comparison to the scale factor $i$, the
parameters $\rho$ or $T$ had relatively little impact on
$\mathbb{U}_{i, \rho, T}$.

We thus propose to use
different values of $\tau$ in Step 3 of Algorithm of Section
\ref{sec:algorithm} and in the within-scale post-processing procedure
of Section \ref{sec:within}. Denoting the former by $\tau_{i, 1}$
and the latter by $\tau_{i, 2}$, we chose $\tau_{i, 1}$ differently
for each $i$ as the $95\%$ quantile, and $\tau_{i, 2}$ as the $97.5\%$ quantile of $\mathbb{U}_{i, \rho, T}$ for given $i$ and $T$ and $\rho$ chosen from the set $\{0, 0.3, 0.6, 0.9\}$ with equal probability. The numerical values of $\mathbb{U}_{i, 0, T}$ when $T=1024$ are summarised in Table \ref{table:tau}.

\begin{table}
\caption{Values of $\tau$ for each scale $i=-1, \ldots, -4$.}
\label{table:tau}
\centering
\begin{tabular}{c c c c c}
\hline
\hline
scale $i$ & $-1$ & $-2$ & $-3$ & $-4$ \\
\hline
$\tau_{i, 1}$ & $0.39$ & $0.46$ & $0.67$ & $0.83$ \\
$\tau_{i, 2}$ & $0.48$ & $0.52$ & $0.75$ & $0.96$ \\
\hline
\end{tabular}
\end{table}

Finally, we discuss the choice of $\is$. We first detect breakpoints in wavelet periodograms at
scales $i=-1, \ldots, -\lfloor \log_2T/3 \rfloor$ and perform the
across-scale post-processing as described in Section
\ref{sec:across}, obtaining the set of breakpoints
$\wh{\mathcal{B}}=\left\{\wh{\nu}_p, \ p=1, \ldots, \wh{B}
\right\}$. Subsequently, for the wavelet periodogram at the next
finest scale, we compute the quantity $\mathbb{V}_p, \ p=1, \ldots,
\wh{B}+1$ as
\[
\mathbb{V}_p=\max_{\nu \in (\wh{\nu}_{p-1}, \wh{\nu}_p)} \left\vert
\frac{
\sqrt{\frac{\wh{\nu}_{p}-\nu}{(\wh{\nu}_{p}-\wh{\nu}_{p-1})\cdot(\nu-\wh{\nu}_{p-1})}}
\sum_{t=\wh{\nu}_{p-1}+1}^{\nu}\iitt
-
\sqrt{\frac{\nu-\wh{\nu}_{p-1}}{(\wh{\nu}_{p}-\wh{\nu}_{p-1})\cdot(\wh{\nu}_p-\nu)}}
\sum_{t=\nu+1}^{\wh{\nu}_{p}}\iitt
}
{\sum_{\wh{\nu}_{p-1}+1}^{\wh{\nu}_p} \iitt/(\wh{\nu}_{p}-\wh{\nu}_{p-1})}\right\vert
\]
where $\wh{\nu}_0=-1$ and $\wh{\nu}_{\wh{B}+1}=T-1$.
Then $\mathbb{V}_p$ is compared to $\tau_{i, 1}\cdot
T^{\theta}\sqrt{\log T}$ to see whether there are any further
breakpoints yet to be detected in $\iitt$ that have not been
included in $\wh{\mathcal{B}}$. (This step is similar to our
within-scale post-processing.) If there is an interval
$[\wh{\nu}_{p-1}+1, \wh{\nu}_{p}]$ where $\mathbb{V}_p$ exceeds
the threshold, $\is$ is updated as $\is := \is + 1$ and the above
procedure is repeated to update $\wh{\mathcal{B}}$ until either no
further changes are made, or $\is \ge \lfloor \log_2 T/2 \rfloor$.

We note that this approach is in line with the theoretical 
consistency of our breakpoint detection procedure; $\mathbb{V}_p$ 
is of the same form as the test statistic and Lemma 6 in the Appendix implies that, 
if there are no more breakpoints to be detected from $\iitt$ for $i<-\is$ 
other than those already chosen ($\wh{\mathcal{B}}$), then $\mathbb{V}_p$ 
does not exceed the threshold, and vice versa by Lemma 5.

\section{Simulation study}
\label{sec:simulation}

In \citet{davis2006}, the performance of the Auto-PARM was assessed
and compared with the Auto-SLEX (\citet{ombao2001}) through simulation
in various settings. The Auto-PARM was shown to be superior to
Auto-SLEX in identifying both dyadic and non-dyadic breakpoints in
piecewise stationary time series. Some examples in the following are
adopted from \citet{davis2006} for the comparative study between our
method and the Auto-PARM, alongside some other new examples. 
We also applied the breakpoint detection method proposed in 
\citet{lavielle2005} to the same simulated processes and, while the
performance was found to be good, it was inferior to both Auto-PARM
and our method for these particular examples, so we do not report these
results.
In the simulations, wavelet periodograms were computed using Haar
wavelets and both post-processing procedures (Section
\ref{sec:within} and Section \ref{sec:across}) followed the
application of the segmentation algorithm. In our examples, $T=1024$
and therefore $I^*$ was set as $3$ at the start of each application
of the algorithm, then updated automatically if necessary, as
described in Section \ref{sec:parameters}.
Simulation outcomes are given in Tables \ref{table:sim:zero}--\ref{table:sim}, where the total number of detected breakpoints are summarised over $100$ simulations.

\begin{figure}[ht]
\begin{minipage}[b]{1\linewidth}
\centering
\includegraphics[width=1\textwidth, height=.25\textheight]{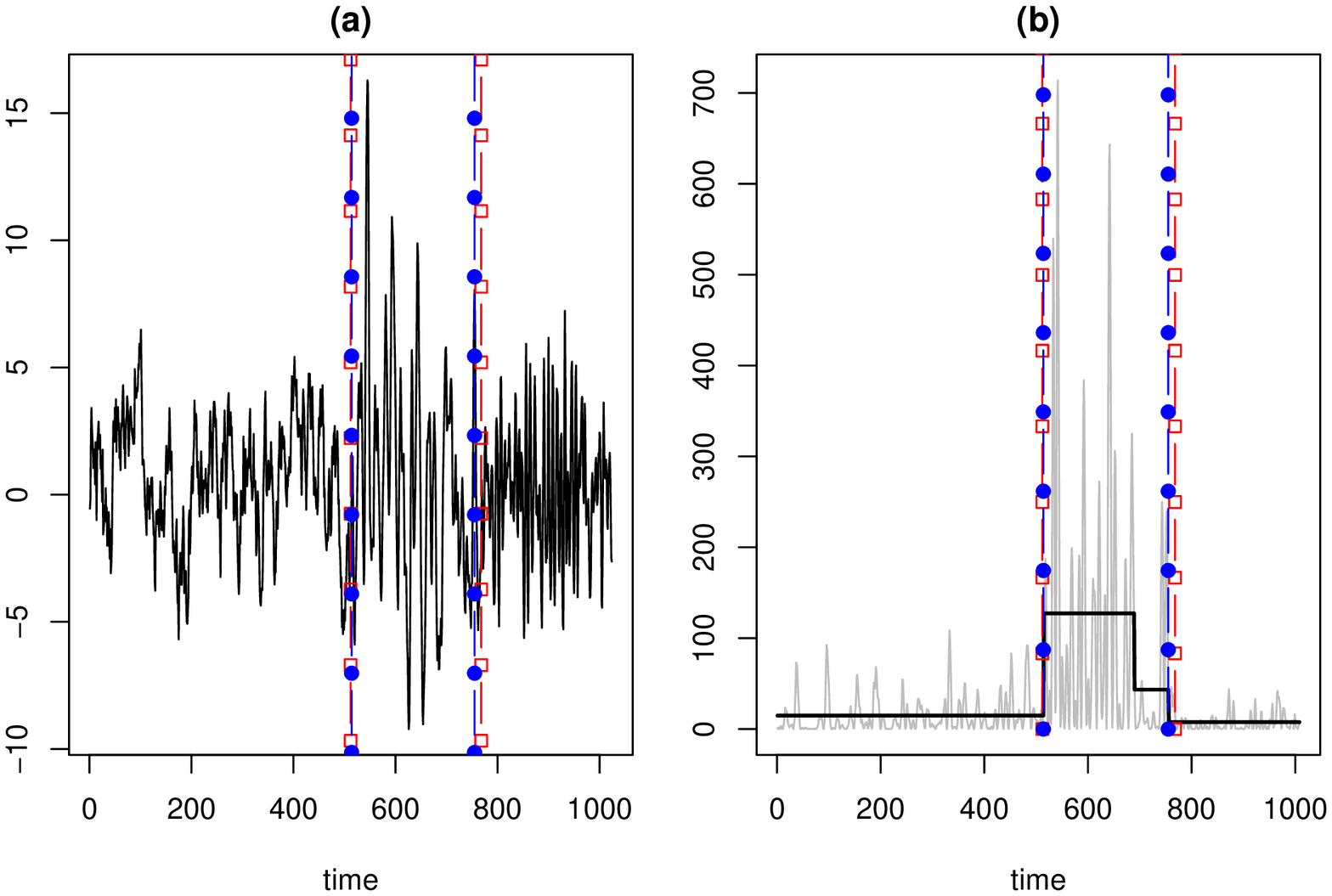} 
\caption{\footnotesize{(a) A realisation of model (\ref{sim:one}) with true (empty square) and detected (filled circle) breakpoints; (b) $\iitt$ at $i=-4$ and the breakpoint detection outcome.}} 
\label{fig:sim:one}
\end{minipage}
\begin{minipage}[b]{1\linewidth}
\centering
\includegraphics[width=1\textwidth, height=.25\textheight]{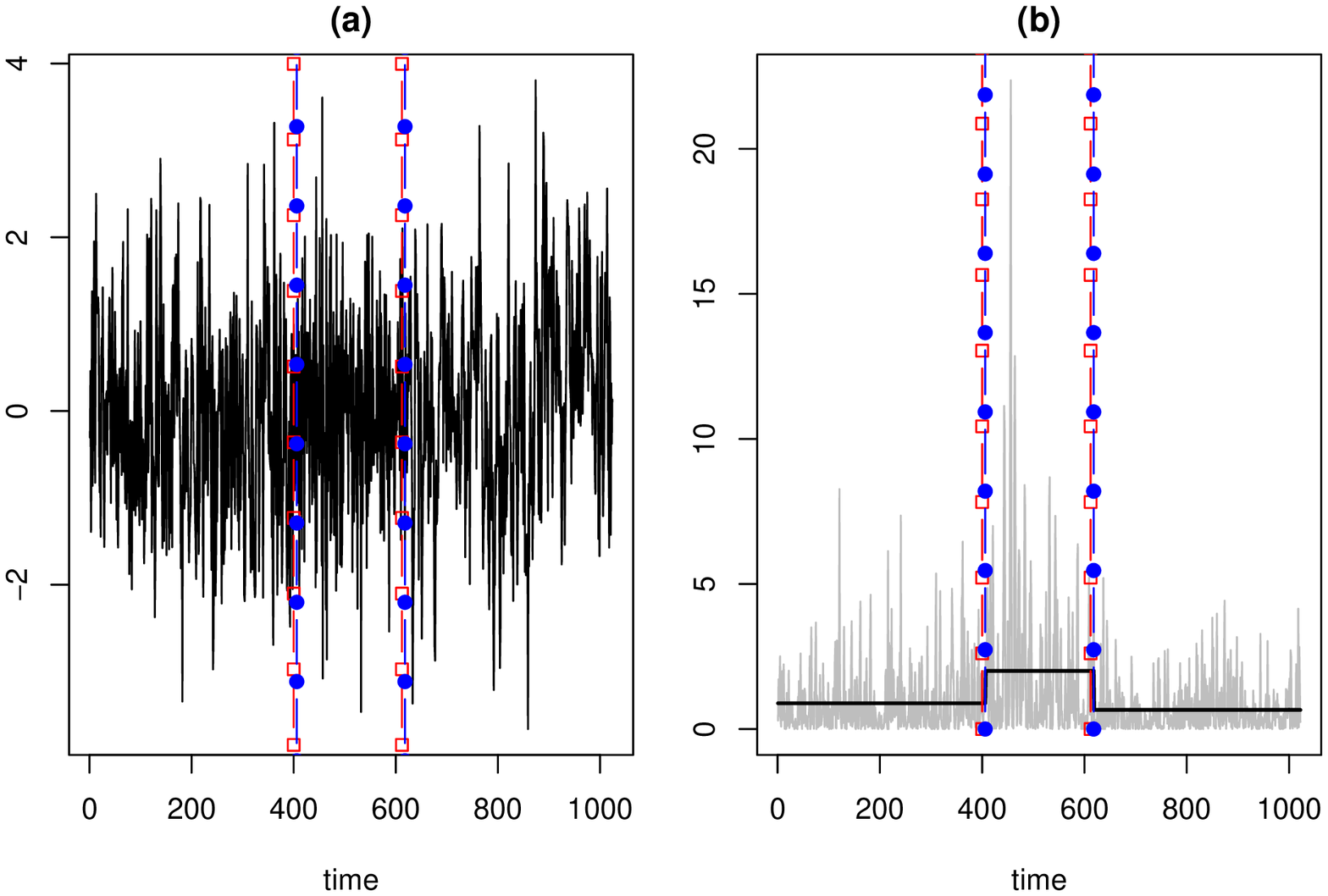}
\caption{\footnotesize{(a) A realisation of model (\ref{sim:two}); (b) $\iitt$ at $i=-1$ and the breakpoint detection outcome.}}
\label{fig:sim:two}
\end{minipage}
\begin{minipage}[b]{1\linewidth}
\centering
\includegraphics[width=1\textwidth, height=.25\textheight]{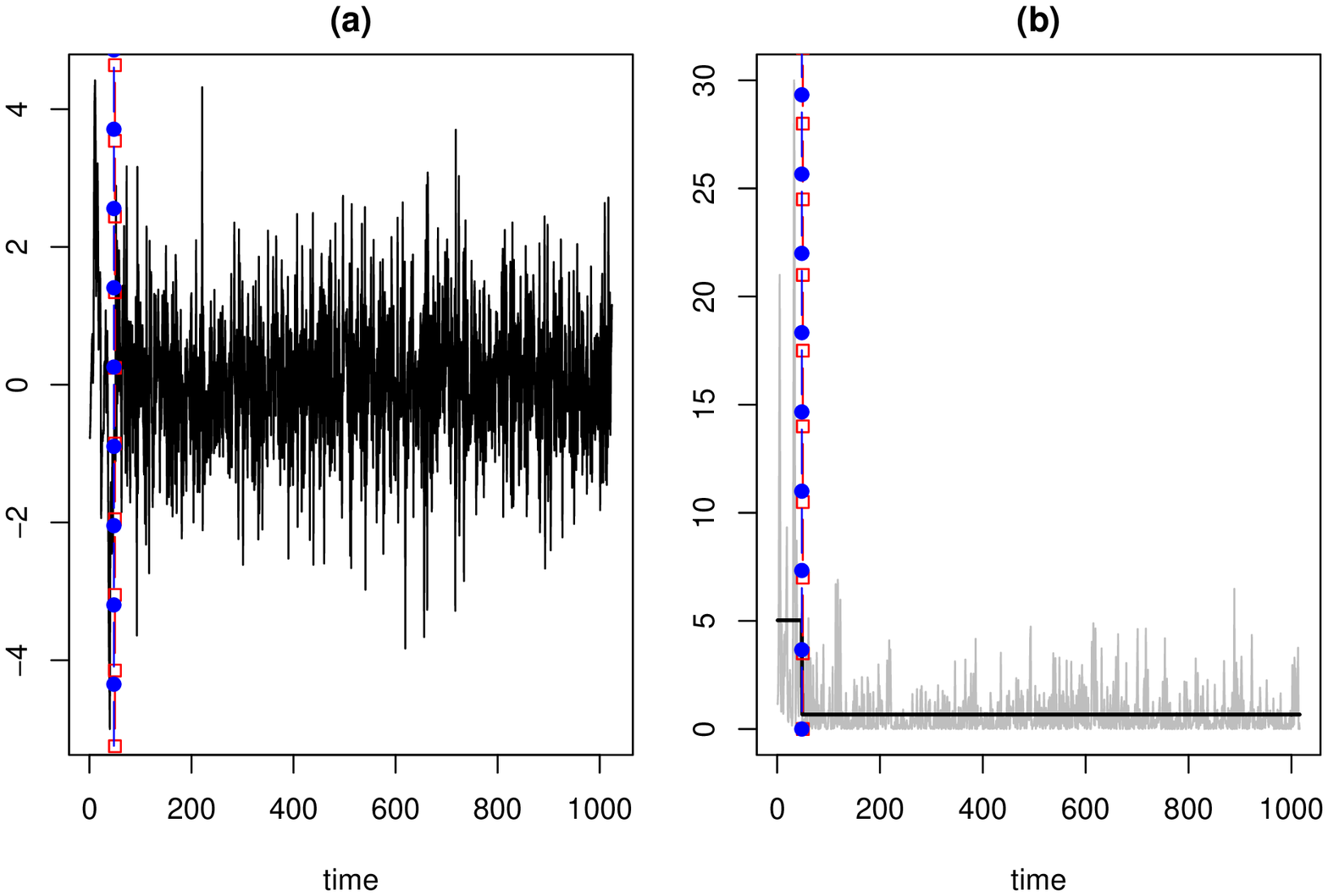}
\caption{\footnotesize{(a) A realisation of model (\ref{sim:three}); (b) $\iitt$ at $i=-3$ and the breakpoint detection outcome.}}
\label{fig:sim:three}
\end{minipage}
\end{figure}

\begin{figure}[ht]
\begin{minipage}[b]{1\linewidth}
\centering
\includegraphics[width=1\textwidth, height=.25\textheight]{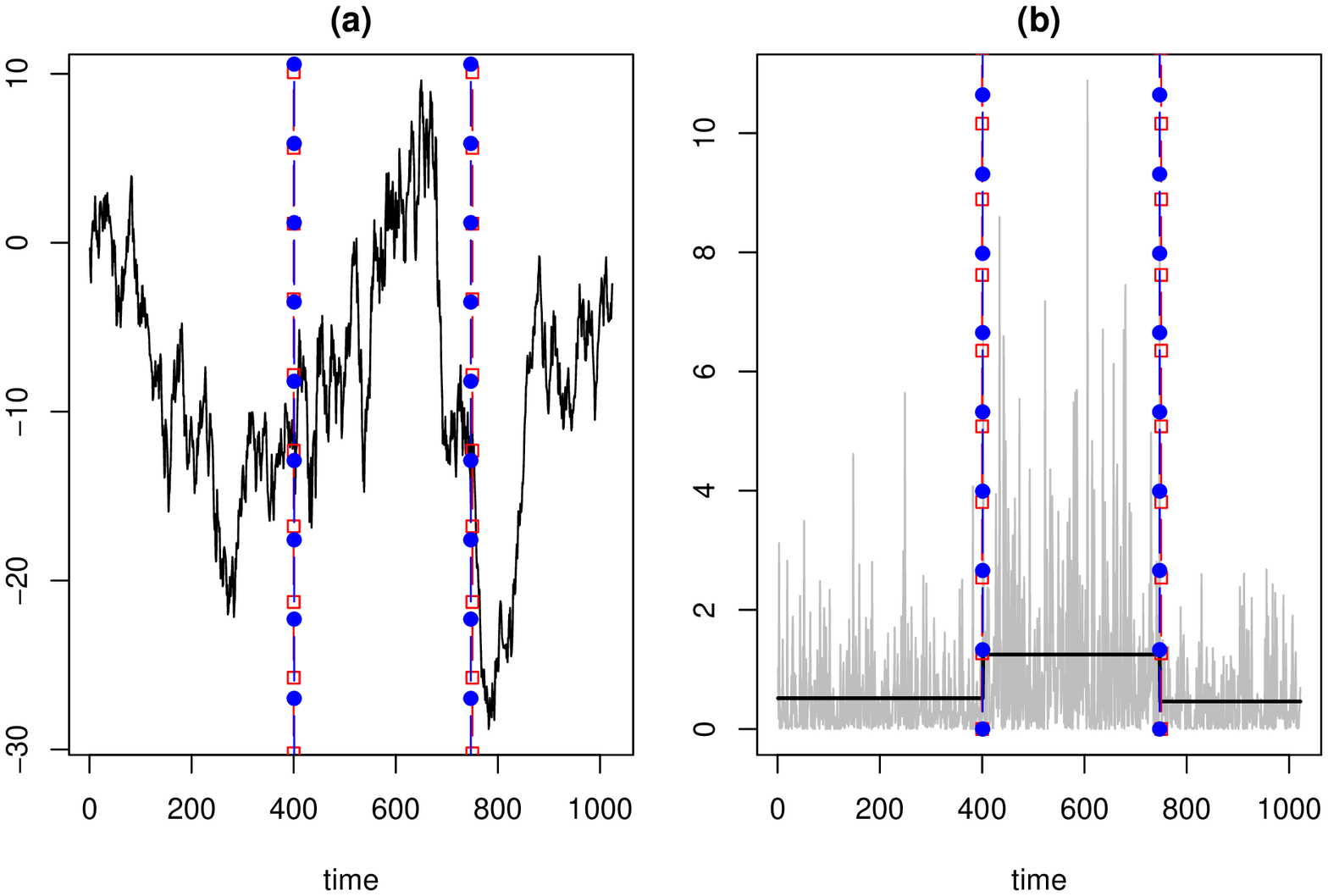}
\caption{\footnotesize{(a) A realisation of model (\ref{sim:four}); (b) $\iitt$ at $i=-1$ and the breakpoint detection outcome.}}
\label{fig:sim:four}
\end{minipage}
\begin{minipage}[b]{1\linewidth}
\centering
\includegraphics[width=1\textwidth, height=.3\textheight]{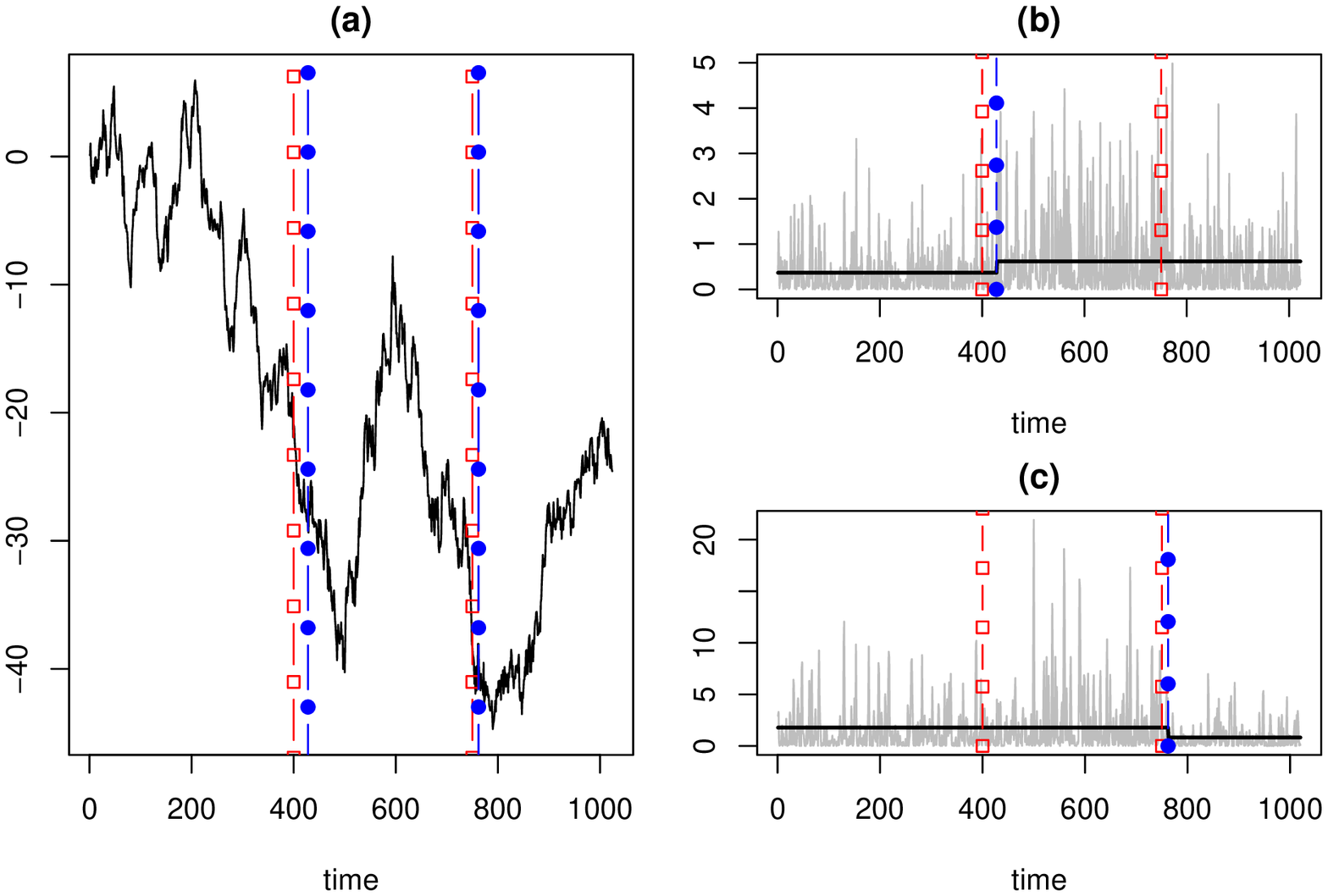}
\caption{\footnotesize{(a) A realisation of model (\ref{sim:five}); (b) $\iitt$ at $i=-1$ and the breakpoint detection outcome; (c) $\iitt$ at $i=-2$ and the breakpoint detection outcome.}}
\label{fig:sim:five}
\end{minipage}
\begin{minipage}[b]{1\linewidth}
\centering
\includegraphics[width=1\textwidth, height=.25\textheight]{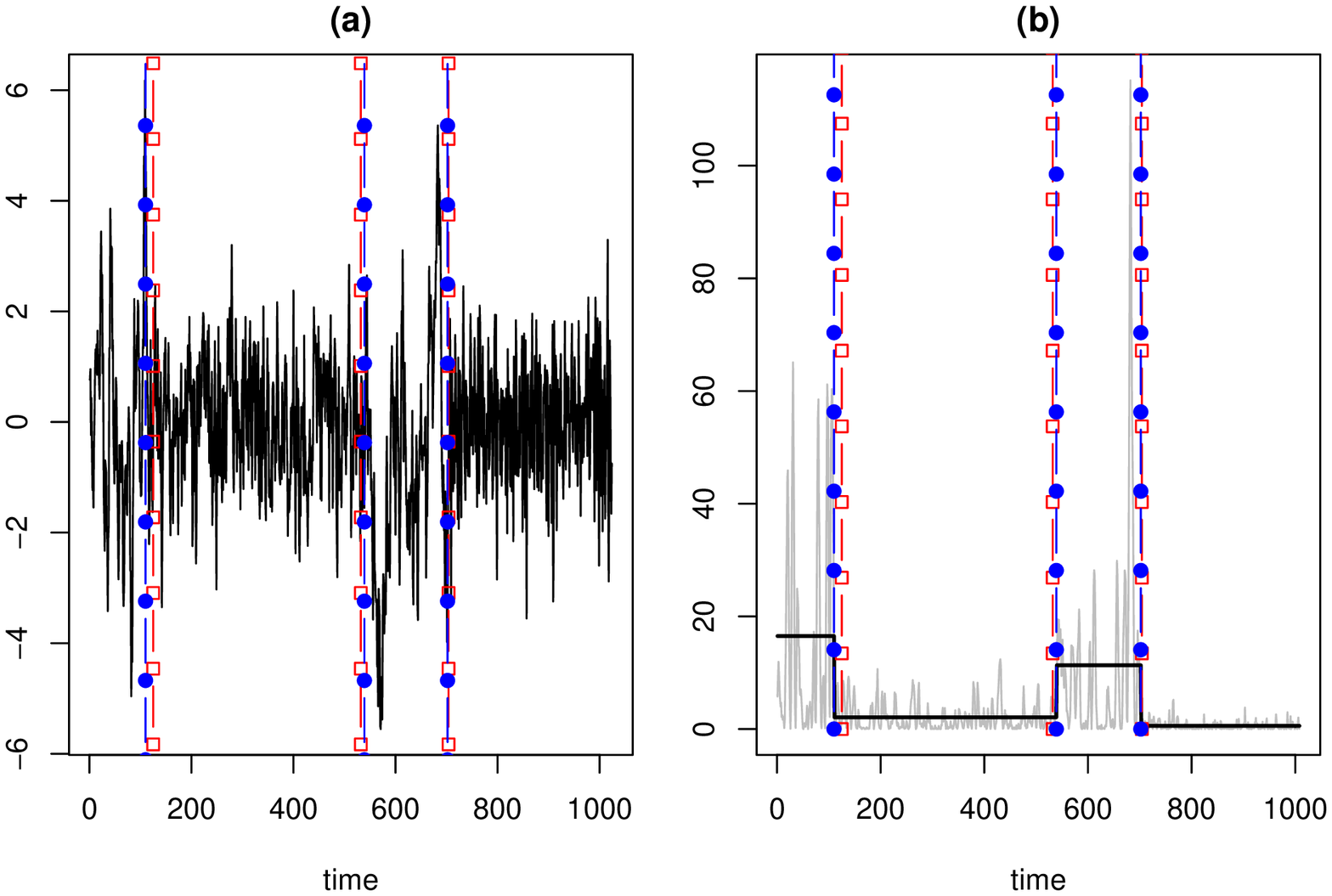}
\caption{\footnotesize{(a) A realisation of model (\ref{sim:seven}); (b) $\iitt$ at $i=-4$ and the breakpoint detection outcome.}}
\label{fig:sim:seven}
\end{minipage}
\end{figure}

\begin{description}
\item[(A) Stationary AR(1) process with no breakpoints] \hfill \\
We consider a stationary AR(1) process,
\begin{eqnarray}
X_t=aX_{t-1}+\ep_t &\mbox{ for } 1 \le t \le 1024,
\label{sim:zero}
\end{eqnarray}
where $\ep_t\sim{\mbox{i.i.d.}}$ $\mathcal{N}(0, 1)$ (as in all subsequent examples unless specified otherwise).
For a range of values of $a$, we summarise the breakpoint detection outcome in Table \ref{table:sim:zero}.

\item[(B) Piecewise stationary AR process with clearly observable changes] \hfill \\
This example is taken from \citet{davis2006}. The target
nonstationary process was generated as
\begin{eqnarray}
X_t=\left\{\begin{array}{ll}
0.9X_{t-1}+\ep_t &\mbox{ for } 1 \le t \le 512, \\
1.68X_{t-1}-0.81X_{t-2}+\ep_t &\mbox{ for } 513 \le t \le 768, \\
1.32X_{t-1}-0.81X_{t-2}+\ep_t &\mbox{ for } 769 \le t \le 1024.
\end{array}\right.
\label{sim:one}
\end{eqnarray}
As seen in Figure \ref{fig:sim:one} (a), there is a clear
difference between the three segments in the model. Figure
\ref{fig:sim:one} (b) shows the wavelet periodogram at scale $-4$
and the estimation results, where the lines with empty squares indicate the true
breakpoints ($\eta_1=512,\eta_2=768$) while the lines with filled circles
indicate the detected ones ($\heta_1=519,\heta_2=764$). Note that
although initially the procedure returned three breakpoints, the
within-sequence post-processing successfully removed the false one. 

\item[(C) Piecewise stationary AR process with less clearly observable changes] \hfill \\
In this example, the piecewise stationary AR model is revisited, but
its breakpoints are less clear-cut, as seen in Figure
\ref{fig:sim:two}.
\begin{eqnarray}
X_t=\left\{\begin{array}{ll}
0.4X_{t-1}+\ep_t & \mbox{ for } 1 \le t \le 400, \\
-0.6X_{t-1}+\ep_t & \mbox{ for } 401 \le t \le 612, \\
0.5X_{t-1}+\ep_t & \mbox{ for } 613 \le t \le 1024 \\
\end{array}\right.
\label{sim:two}
\end{eqnarray}
Figure \ref{fig:sim:two} (b) shows the wavelet periodogram at scale $-1$ for the realisation in the left panel with its breakpoint estimates ($\heta_1=403,\heta_2=622$).
Both procedures achieved good performance.

\item[(D) Piecewise stationary AR process with a short segment] \hfill \\
This example is again from \citet{davis2006}. A single breakpoint occurs and one segment is much shorter than the other.
\begin{eqnarray}
X_t=\left\{\begin{array}{ll}
0.75X_{t-1}+\ep_t & \mbox{ for } 1 \le t \le 50, \\
-0.5X_{t-1}+\ep_t & \mbox{ for } 51 \le t \le 1024. \\
\end{array}\right.
\label{sim:three}
\end{eqnarray}
A typical realisation of (\ref{sim:three}), its wavelet periodogram at scale $-3$, and the estimation outcome are shown in Figure \ref{fig:sim:three},
where the jump at $\eta_1=50$ was identified as $\heta_1=49$.
Even though one segment is substantially shorter than the other, our procedure was able to detect exactly one breakpoint in $97\%$ of the cases and underestimation did not occur even when it failed to detect exactly one.

\item[(E) Piecewise stationary near-unit-root process with changing variance] \hfill \\
Financial time series, such as stock indices, individual share or commodity prices, or currency exchange rates are, 
for such purposes as pricing of derivative instruments, often modelled by
a random walk with a time-varying variance. We generated a piecewise stationary, near-unit-root example following (\ref{sim:four}), where the variance has two breakpoints over time and the AR parameter remains constant and very close to 1; 
a typical realisation is given in Figure \ref{fig:sim:four} (a).
Note that, within each stationary segment, the process can be seen as a special case of 
the near-unit-root process of \citet{phillips1988}.

\begin{eqnarray}
X_t=\left\{\begin{array}{lll}
0.999X_{t-1}+\ep_t, & \ep_t\sim\mathcal{N}(0, 1) & \mbox{ for } 1 \le t \le 400, \\
0.999X_{t-1}+\ep_t, & \ep_t\sim\mathcal{N}(0, 1.5^2) & \mbox{ for } 401 \le t \le 750, \\
0.999X_{t-1}+\ep_t, & \ep_t\sim\mathcal{N}(0, 1) & \mbox{ for } 751 \le t \le 1024. \\
\end{array}\right.
\label{sim:four}
\end{eqnarray}
Recall that the Auto-PARM is designed to find the ``best'' combination of the total number 
and locations of breakpoints, and adopts a genetic algorithm to traverse the vast parameter space.
However, due to the stochastic nature of the algorithm, it occasionally fails to return consistent estimates.
This instability was emphasised here, with each run often returning different breakpoints.
For one typical realisation, it detected $t=21$ and $797$ as breakpoints, and then only $t=741$ in the next run on the same sample path.
Overall, the performance of Auto-PARM leaves much to be desired for this particular example, whereas our method performed well, though this is not a criticism of Auto-PARM in general, as it performed well in other examples.
Note that it was at scale $-1$ of the wavelet periodogram that both breakpoints were consistently identified the most frequently.
The computation of the wavelet periodogram at scale $-1$ with Haar wavelets is a differencing operation and naturally ``whitens'' the near-unit-root process (\ref{sim:four}), clearly revealing any changes of variance in the sequence.

\item[(F) Piecewise stationary AR process with high autocorrelation] \hfill \\
The features of this AR model are a high degree of autocorrelation
and less obvious breakpoints compared to previous examples. A typical realisation 
is shown in Figure \ref{fig:sim:five} (a).
\begin{eqnarray}
X_t=\left\{\begin{array}{lll}
1.399X_{t-1}-0.4X_{t-1}+\ep_t, & \ep_t\sim\mathcal{N}(0, 0.8^2) & \mbox{ for } 1 \le t \le 400, \\
0.999X_{t-1}+\ep_t, & \ep_t\sim\mathcal{N}(0, 1.2^2) & \mbox{ for } 401 \le t \le 750, \\
0.699X_{t-1}+0.3X_{t-1}+\ep_t, & \ep_t\sim\mathcal{N}(0, 1) & \mbox{ for } 751 \le t \le 1024. \\
\end{array}\right.
\label{sim:five}
\end{eqnarray}
Again, the instability of Auto-PARM was notable here, with the second breakpoint at $t=750$ often left undetected.
Our procedure correctly identified both breakpoints in $84\%$ of the cases.

\item[(G) Piecewise stationary ARMA$(1, 1)$ process] \hfill \\
We generated piecewise stationary ARMA processes as
\begin{eqnarray}
X_t=\left\{\begin{array}{lll}
0.7X_{t-1}+\ep_t+0.6\ep_{t-1} & \mbox{ for } 1 \le t \le 125, \\
0.3X_{t-1}+\ep_t+0.3\ep_{t-1} & \mbox{ for } 126 \le t \le 532, \\
0.9X_{t-1}+\ep_t & \mbox{ for } 533 \le t \le 704, \\
0.1X_{t-1}+\ep_t-0.5\ep_{t-1} & \mbox{ for } 705 \le t \le 1024.
\end{array}\right.
\label{sim:seven}
\end{eqnarray}
As illustrated in Figure \ref{fig:sim:seven} (a), the first breakpoint $
t=125$ is less apparent than the other two. Auto-PARM often left this breakpoint
undetected, while our procedure found all three in $76\%$ of cases. 
We note that it was scale $i=-4$ at which $t=125$ was detected most frequently by our procedure.
With a time series of length $T=1024$, default scales provided by our algorithm are 
$i=-1, -2, -3$, and this example demonstrates the effectiveness of the 
updating procedure for $\is$ described in Section \ref{sec:parameters}. That is, 
after completing the examination of $\iitt$ for $i=-1, -2, -3$, our procedure checked 
if there were more breakpoints to be detected from $\iitt$ for the next scale $i=-4$ 
and, as it was the case, updated $\is$ to $4$. Figure \ref{fig:sim:seven} (b) shows 
the wavelet periodogram at scale $-4$ for the time series example in the left panel.
\end{description}

\begin{table}
\caption{Summary of breakpoint detection from Simulation (A): our method (CF) and Auto-PARM (AP). Results over $100$ simulations.}
\label{table:sim:zero}
\centering
\begin{tabular}{ r  c  c | c  c | c  c | c  c | c  c | c  c }
\hline
\hline
\multicolumn{13}{c}{number of breakpoints} \\
a & \multicolumn{2}{c}{0.7}
& \multicolumn{2}{c}{0.4}
& \multicolumn{2}{c}{0.1} 
& \multicolumn{2}{c}{-0.1} 
& \multicolumn{2}{c}{-0.4} 
& \multicolumn{2}{c}{-0.7} \\
& CF & AP &
CF & AP &
CF & AP &
CF & AP &
CF & AP &
CF & AP \\
\hline
0 &
\textbf{100} & \textbf{100} &
\textbf{100} & \textbf{100} &
\textbf{100} & \textbf{100} &
\textbf{99} & \textbf{100} &
\textbf{99} & \textbf{100} &
\textbf{94} & \textbf{100}
\\
1 &
0 & 0 &
0 & 0 &
0 & 0 &
1 & 0 &
1 & 0 &
5 & 0 
\\
$\ge$ 2 &
0 & 0 &
0 & 0 &
0 & 0 &
0 & 0 &
0 & 0 &
1 & 0
\\
total &
100 & 100 &
100 & 100 &
100 & 100 &
100 & 100 &
100 & 100 &
100 & 100
\\
\hline
\end{tabular}
\end{table}

\begin{table}
\caption{Summary of breakpoint detection from simulations: our method (CF) and Auto-PARM (AP). Results over $100$ simulations.}
\label{table:sim}
\centering
\begin{tabular}{ r  c  c | c  c | c  c | c  c | c  c | c  c}
\hline
\hline
\multicolumn{13}{c}{number of breakpoints} \\
& \multicolumn{2}{c}{model (B)}
& \multicolumn{2}{c}{model (C)}
& \multicolumn{2}{c}{model (D)}
& \multicolumn{2}{c}{model (E)} 
& \multicolumn{2}{c}{model (F)} 
& \multicolumn{2}{c}{model (G)} \\
& CF & AP &
CF & AP &
CF & AP &
CF & AP &
CF & AP &
CF & AP \\
\hline
0 &
0 & 0 &
0 & 0 &
0 & 0 &
1 & 42 &
1 & 20 &
0 & 0
\\
1 &
0 & 0 &
0 & 0 &
\textbf{97} & \textbf{100} &
0 & 31 &
14 & 68 &
1 & 16
\\
2 &
\textbf{93} & \textbf{99} &
\textbf{96} & \textbf{100} &
3 & 0 &
\textbf{97} & \textbf{16} &
\textbf{84} & \textbf{7} &
6 & 55
\\
3 &
4 & 1 &
3 & 0 &
0 & 0 &
2 & 9 &
1 & 3 &
\textbf{76} & \textbf{29} 
\\
4 &
3 & 0 &
1 & 0 &
0 & 0 &
0 & 0 &
0 & 1 &
17 & 0
\\
5 &
0 & 0 &
0 & 0 &
0 & 0 &
0 & 2 &
0 & 1 &
0 & 0
\\
total &
100 & 100 &
100 & 100 &
100 & 100 &
100 & 100 &
100 & 100 &
100 & 100
\\
\hline
\end{tabular}
\end{table}

\section{U.S stock market data analysis}
\label{sec:real}

Many authors, including \citet{starica2005}, argue in favour of nonstationary modelling of financial returns.
In this analysis, we consider the Dow Jones Industrial Average index and regard it as a process
with an extremely high degree of autocorrelation (such as in the near-unit-root model of
\citet{phillips1988}) and a time-varying variance, similar to the simulation model in Section 
\ref{sec:simulation} (E).

\begin{figure}[ht]
\begin{minipage}[b]{1\linewidth}
\centering
\includegraphics[width=1\textwidth, height=.3\textheight]{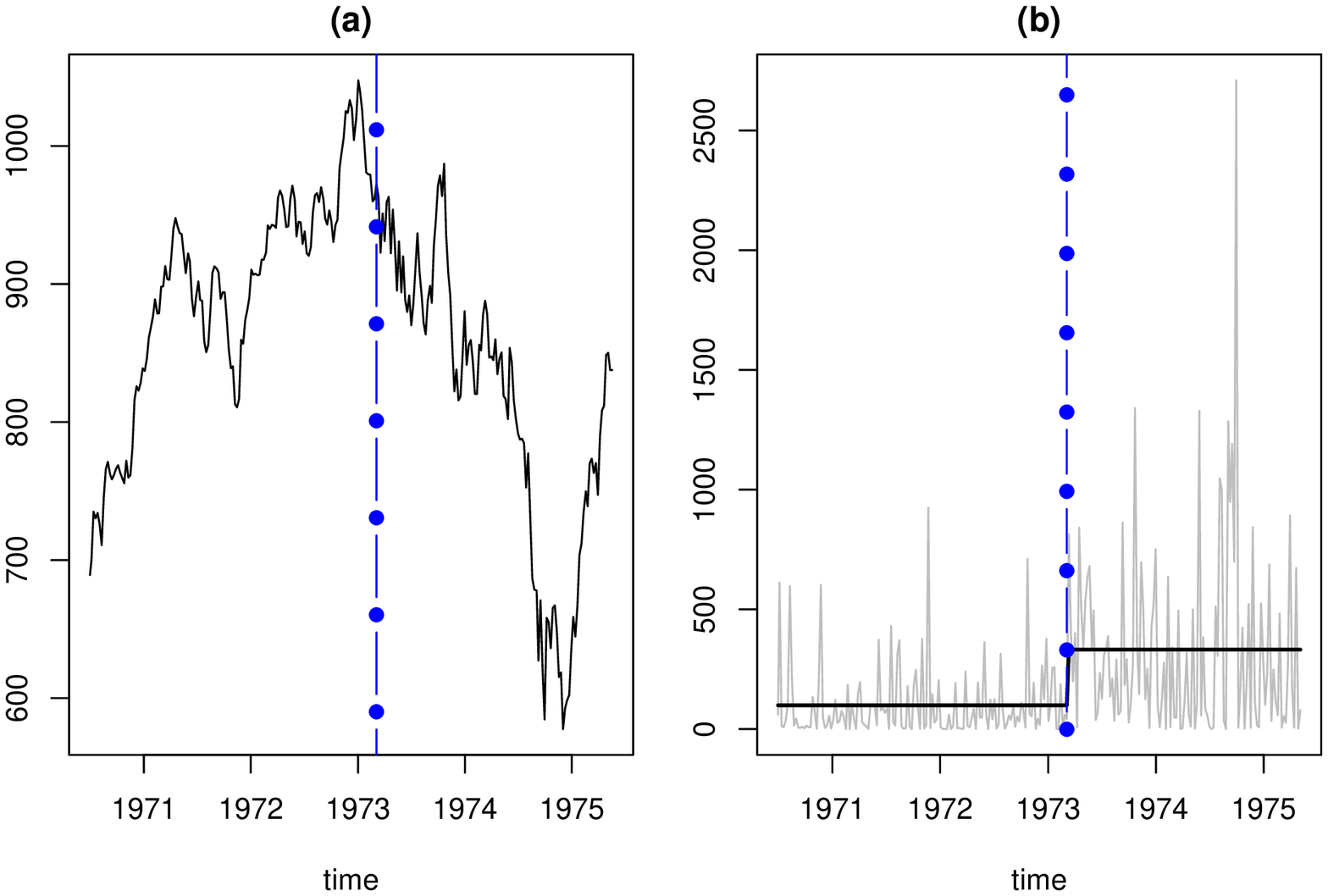}
\caption{\footnotesize{(a) Weekly average values of the Dow Jones IA index (July 1970--May 1975); (b) Wavelet periodogram at scale $-1$ and the breakpoint detection outcome.}}
\label{fig:real:one}
\end{minipage}
\vspace{0.5cm}
\begin{minipage}[b]{1\linewidth}
\centering
\includegraphics[width=1\textwidth, height=.3\textheight]{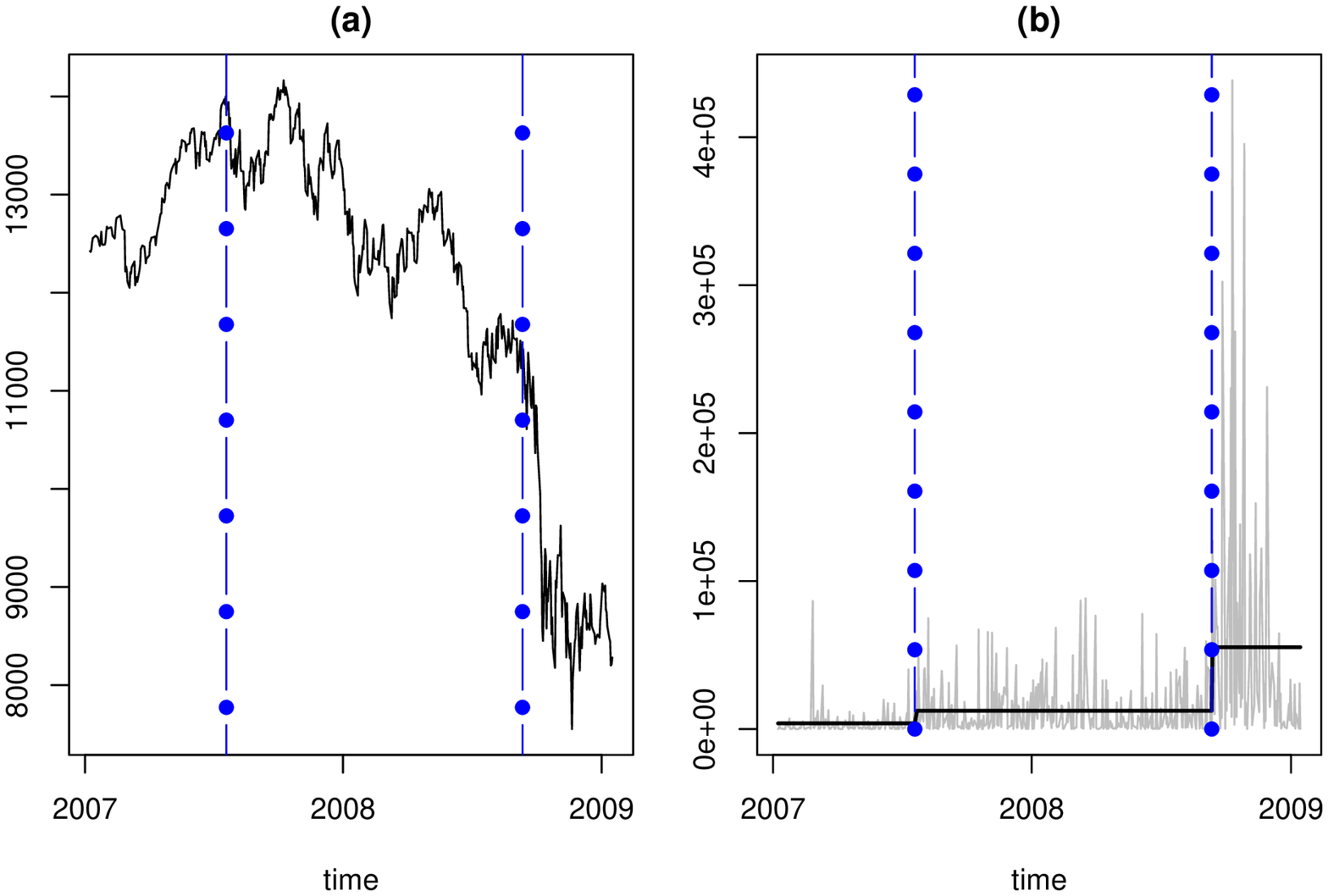}
\caption{\footnotesize{(a) Daily average values of the Dow Jones IA index (Jan 2007--Jan 2009); (b) Wavelet periodogram at scale $-1$ and the breakpoint detection outcome.}}
\label{fig:real:two}
\end{minipage}
\end{figure}

\begin{description}
\item[(A) Dow Jones weekly closing values 1970--1975] \hfill \\
The time series of weekly closing values of the Dow Jones Industrial Average index between July 1971 and August 1974 was studied in \citet{hsu1979} and revisited in \citet{chen1997}.
Historical data are available on  \url{www.google.com/finance/historical?q=INDEXDJX:.DJI}, where daily and weekly prices can be extracted for any time period.
Both papers concluded that there was a change in the variance of the index around the third week of March 1973.
For ease of computation of the wavelet periodogram, we chose the same weekly index between 1 July 1970 and 19 May 1975 so that the data size was $T=256$ with the above-mentioned time period was contained in this interval. The third week of March 1973 corresponds to $t=141$ and our procedure detected $\heta=142$ as a breakpoint, as illustrated in Figure \ref{fig:real:one}.
The Auto-PARM did not return any breakpoint, while the segmentation procedure proposed in \citet{lavielle2005}, when applied to the 
log-returns ($\log(X_t/X_{t-1})$) of the data rather than the data $X_t$ themselves, returned $t=141$ as a breakpoint, 
which is very close to $\heta$.

\item[(B) Dow Jones daily closing values 2007--2009] \hfill \\
We further investigated more recent {\em daily} data from the same source, between 8 January 2007 and 16 January 2009.
Over this period, the global financial market experienced one of the worst crises in history.
Our breakpoint detection algorithm found two breakpoints (see Figure \ref{fig:real:two}), one in the last week of July 2007 ($\heta_1=135$), and the other in mid-September 2008 ($\heta_2=424$).
The Auto-PARM returned three breakpoints on average, although the estimated breakpoints were unstable as in Section \ref{sec:simulation} (E):
$t=35, 426$ and $488$ were detected most often as breakpoints, while $t=100$ and $t=140$ were detected in place of $t=35$ on other occasions.
The segmentation procedure from \citet{lavielle2005},
when applied to the 
log-returns ($\log(X_t/X_{t-1})$) of the data rather than the data $X_t$ themselves,
detected $t=127$ and $424$ as breakpoints, which are very close to $\heta_1$ and $\heta_2$.
The first breakpoint coincided with the outbreak of the worldwide ``credit crunch'' as subprime mortgage-backed securities were discovered in portfolios of banks and hedge funds around the world.
The second breakpoint coincided with the bankruptcy of Lehman Brothers, a major financial services firm, an event that brought even more volatility to the market.
Evidence supporting our breakpoint detection outcome is the TED spread
(available from \url{http://www.bloomberg.com/apps/quote?ticker=.tedsp:ind}), an 
indicator of perceived credit risk in the general economy; it spiked up in 
late July 2007, remained volatile for a year, then spiked even higher in September 2008.
These movements coincide almost exactly with our detected breakpoints.
\end{description}

\section*{Acknowledgements}

The authors would like to thank Rainer von Sachs for his interesting comments to this work.
Also we are grateful to the Editor, an associate editor and two referees for their stimulating reports that led to a significant improvement of the paper.

\appendix
\section{The proof of Theorem 1}
\label{sec:proof:one}

The consistency of our algorithm is first proved for the sequence below,
\begin{eqnarray}
\tystt=\sst\cdot\zstt, \ t=0, \ldots, T-1.
\label{bs:generic}
\end{eqnarray}
Note that unlike in (3), the above model features the
true piecewise constant $\sst$. Denote $n=e-s+1$ and define
\begin{eqnarray*}
\Y^b_{s, e}&=&\sumsb \tystt- \sumbe \tystt.
\end{eqnarray*}
$\bS^b_{s, e}$ and $\bhS^b_{s, e}$ are defined similarly, replacing $\tystt$ with 
$\sst$ and $\sstt$, respectively. Note that the above are simply inner products of the respective
sequences and a vector whose support starts at $s$, is constant and
positive until $b$, then constant negative until $e$, and normalised
such that it sums to zero and sums to one when squared. Let $s$, $e$
satisfy $ \eta_{p_0} \le s < \eta_{p_0+1} < \ldots < \eta_{p_0+q} <
e \le \eta_{p_0+q+1} $ for $0 \le p_0 \le B-q$, which will always be
the case at all stages of the algorithm. In Lemmas 1--5 below, we
impose at least one of following conditions:
\begin{eqnarray}
s< \eta_{p_0+r}-C\delt < \eta_{p_0+r}+C\delt < e \mbox{ for some } 1 \le r \le q,
\label{lem:cond:one} \\
\{(\eta_{p_0+1}-s)\wedge(s-\eta_{p_0})\} \vee \{(\eta_{p_0+q+1}-e)\wedge(e-\eta_{p_0+q})\} \le \ept,
\label{lem:cond:two}
\end{eqnarray}
where $\wedge$ and $\vee$ are the minimum and maximum operators,
respectively. We remark that both conditions (\ref{lem:cond:one})
and (\ref{lem:cond:two}) hold throughout the algorithm for all those
segments starting at $s$ and ending at $e$ which contain previously
undetected breakpoints. As Lemma 6 concerns the case when all
breakpoint have already been detected, it does not use either of
these conditions.

The proof of the theorem is constructed as follows. Lemma \ref{lem:one} is used in the proof of Lemma \ref{lem:two}, 
which in turn is used alongside Lemma \ref{lem:three} in the proof of Lemma \ref{lem:four}. 
From the result of Lemma \ref{lem:four}, we derive Lemma \ref{lem:five} and finally, 
Lemmas \ref{lem:five} and \ref{lem:six} are used to prove Theorem 1.

\begin{lem}
\label{lem:one} Let $s$ and $e$ satisfy (\ref{lem:cond:one}), then there exists $1 \le r^* \le q$ such that
\begin{eqnarray}
\label{lem:one:eq}
\left|\bS^{\eta_{p_0+r^*}}_{s, e}\right| =\max_{s<t<e}|\bS^{t}_{s, e}|
\ge C\delt/\sqrt{T}. 
\end{eqnarray}
\end{lem}
\noindent \textbf{Proof.}
The equality in (\ref{lem:one:eq}) is proved by Lemmas 2.2 and 2.3 of Venkatraman (1993).
For the inequality part, we note that in the case of a single breakpoint in $\sigma^2(z)$, $r$ in (\ref{lem:cond:one}) coincides with $r^*$ and we can use the constancy of $\sigma^2(z)$ to the left and to the right of the breakpoint to show that
\begin{eqnarray*}
\left|\bS^{\eta_{p_0+r}}_{s, e}\right|
=\left|\frac{\sqrt{\eta_{p_0+r}-s+1}\sqrt{e-\eta_{p_0+r}}}{\sqrt{n}}
\left(\sigma^2(\eta_{p_0+r}/T)-\sigma^2((\eta_{p_0+r}+1)/T)\right)\right|,
\end{eqnarray*}
which is bounded from below by $C\delt/\sqrt{T}$. 
In the case of multiple breakpoints, we remark that for any $r$ satisfying (\ref{lem:cond:one}), the above order remains the same and thus (\ref{lem:one:eq}) follows. \hfill$\square$

\begin{lem}
\label{lem:two} Suppose (\ref{lem:cond:one}) holds, and 
let $\eta \equiv \eta_{p_0+r} \in [s, e]$ for some $r\in\{1, \ldots, q\}$, denote a true change-point.
Then there exists $c_0\in(0, \infty)$ such that for $b$ satisfying
$|\bS_{s, e}^b| < |\bS_{s, e}^\eta|$ and $|\eta-b| \ge c_0\ept$ with $\ept = T^{5/2-2\Theta}\log\,T$,
we have $|\bS_{s, e}^\eta| - |\bS_{s, e}^b| \ge 2\log\,T$.
\end{lem}
\noindent \textbf{Proof.}
Let both $\bS_{s, e}^\eta$, $\bS_{s, e}^b$ $\ge 0$ without loss of generality.

The proof follows directly from the proof of Lemma 2.6 in \cite{venkatraman1993}.
We only consider Case 2 of Lemma 2.6, 
since adapting the proof of Case 1 (when there is a single change-point within $[s, e]$) 
to that of the current lemma takes analogous arguments.

Using the notations therein, it is shown that the term $E_{1l}$ is dominant over $E_{2l}$ and $E_{3l}$
in $\bS_{s, e}^\eta - \bS_{s, e}^b$, where $l = c_0\ept$. 
Noting further that $i=\eta-s+1$, $h=\delta_T$, $j=e-\eta-h$ and $a = \sum_{t=s}^\eta \sigma^2(t/T) - (e-s+1)^{-1}\sum_{t=s}^e\sigma^2(t/T)$,
and that $h \ge 2l$,
\begin{eqnarray*}
E_{1l} &=& \frac{la\sqrt{i+j+h}}{\sqrt{i}\sqrt{j+h}} \cdot \frac{h-l}{\sqrt{i+l}\sqrt{j+h-l}\{\sqrt{(i+l)(j+h-l)}+\sqrt{i(j+l)}\}}
\\
&\ge& \bS_{s, e}^\eta \cdot C\ept\delta_T T^{-2} \ge 2\log\,T
\end{eqnarray*}
for large $T$. 
\hfill$\square$

\begin{lem}
\label{lem:three} $\left|\Y^b_{s, e}-\bS^b_{s, e}\right| \le \log T$ with
probability converging to $1$ with $T$ uniformly over $(s, b, e)\in\mathcal{D}$, where, for $c\in(0, \infty)$, 
\[
\mathcal{D}:=\left\{1 \le s <b<e\le T; \ e-s+1 \ge C\delt, \ \max\left\{\sqrt{\frac{b-s+1}{e-b}},
\sqrt{\frac{e-b}{s-b+1}}\right\} \le c\right\}.
\]
\end{lem}
\noindent \textbf{Proof.}
We need to show that
\begin{eqnarray}
\mathbf{P}\left(\max_{(s, b,
e)\in\mathcal{D}}\frac{1}{\sqrt{n}}\left\vert\sum_{t=s}^e
\sst(\zstt-1)\cdot c_t\right\vert>\log T\right) \longrightarrow 0,
\label{lem:three:one}
\end{eqnarray}
where $c_t=\sqrt{e-b}/\sqrt{b-s+1}$ for
$t\in[s, b]$ and $c_t=\sqrt{b-s+1}/\sqrt{e-b}$ otherwise. Let
$\{U_t\}_{t=s}^e$ be i.i.d. standard normal variables,
$\mathbf{V}=(v_{i, j})_{i, j=1}^n$ with $v_{i, j}=\corr\left(Z_{i,
T}, Z_{j, T}\right)$, and $\mathbf{W}=(w_{i, j})_{i, j=1}^n$ be a
diagonal matrix with $w_{i, i}=\sst\cdot c_t$ where $i=t-s+1$. By
standard results (see e.g. \citet{jk1970}, page 151), showing
(\ref{lem:three:one}) is equivalent to showing that
$\left\vert\sum_{t=s}^e \lam_{t-s+1}(U_t^2-1)\right\vert$ is bounded
by $\sqrt{n}\log T$ with probability converging to 1, where $\lam_i$
are eigenvalues of the matrix $\mathbf{VW}$. Due to the Gaussianity
of $U_t$, $\lam_{t-s+1}(U_t^2-1)$ satisfy the Cram\'{e}r's
condition, i.e., there exists a constant $C>0$ such that
\[
\E\left\vert\lam_{t-s+1}(U_t^2-1)\right\vert^p\le C^{p-2}p!
\E\left\vert\lam_{t-s+1}(U_t^2-1)\right\vert^2, \ p=3, 4, \ldots.
\]
Therefore we can apply Bernstein's inequality \citep{bosq1998} and obtain
\[
\mathbf{P}\left(\left\vert\sum_{t=s}^e
\sst(\zstt-1)\cdot c_t\right\vert>\sqrt{n}\log T\right) \le 2\exp\left(-\frac{n\log^2T}{4\sum_{i=1}^n\lam_i^2+2\max_i|\lam_i|C\sqrt{n}\log T}\right).
\]
Note that 
$
\sum_{i=1}^n\lam_i^2=\tr\left(\mathbf{VW}\right)^2\le
c^2\max_z\sig^4(z)n\rho_{\infty}^2.
$ 
Also it follows that
$\max_i|\lam_i| \le c\max_z$ $\sig^2(z)\Vert \mathbf{V} \Vert$ where
$\Vert\cdot\Vert$ denotes the spectral norm of a matrix, and $\Vert
\mathbf{V}\Vert\le\rho^1_{\infty}$ since $\mathbf{V}$ is
non-negative definite. 
Then (\ref{lem:three:one}) is bounded by
\begin{eqnarray*}
&&\sum_{(s, b, e)\in\mathcal{D}} 2\exp\left(-\frac{n\log^2T}{4c^2\max_z\sig^4(z)n\rho_{\infty}^2+2c\max_z\sig^2(z)\sqrt{n}\log
T\rho^1_{\infty}}\right) \\
&&\le 2T^3\exp\left(-C\log^2T\right) \rightarrow 0,
\end{eqnarray*}
as $\rho_{\infty}^p\le C2^{\is}$, which can be made to be of order
$\log T$, since the only requirement on $I^*$ is that it converges
to infinity but no particular speed is required. Thus the lemma follows. \hfill$\square$

\begin{lem}
\label{lem:four}
Assume (\ref{lem:cond:one}) and (\ref{lem:cond:two}).
For $b=\arg\max_{s<t<e}|\Y^t_{s, e}|$, there exists $1\le r \le q$ such that
$|b-\eta_{p_0+r}|\le C\ept$ for a large $T$.
\end{lem}
\noindent \textbf{Proof.}
Let $\bS_{s, e}=\max_{s<t<e}|\bS^t_{s, e}|$.
From Lemma \ref{lem:three},
$\Y^b_{s, e}\ge\bS_{s, e}-\log T$ and $\bS^b_{s, e}\ge\Y^{b}_{s, e}-\log T$,
hence $\bS^b_{s, e}\ge\bS_{s, e}-2\log T$.
Assume that $|b-\eta_{p_0+r}| > C\ept$ for any $r$.
From Lemma 2.2 in Venkatraman (1993), $\bS^t_{s, e}$ is either monotonic or decreasing and then increasing on $[\eta_{p_0+r}, \eta_{p_0+r+1}]$ and
$\bS^{\eta_{p_0+r}}_{s, e}\vee\bS^{\eta_{p_0+r+1}}_{s, e}>\bS^{b}_{s, e}$.
Suppose that $\bS^t_{s, e}$ is decreasing and then increasing on the interval.
Then from Lemma \ref{lem:two}, we have $b'=\eta_{p_0+r}+C\ept$ satisfying $\bS^{\eta_{p_0+r}}_{s, e}-2\log T\ge\bS^{b'}_{s, e}$.
Since $\bS^t_{s, e}$ is locally increasing at $t=b$ (for $\bS^{b}_{s, e}>\bS^{b'}_{s, e}$), 
we have $\bS^{\eta_{p_0+r+1}}_{s, e}>\bS^{b}_{s, e}$ and
there will again be a $b''=\eta_{p_0+r+1}-C\ept$ satisfying
$\bS^{\eta_{p_0+r}}_{s, e}-2\log T\ge\bS^{b''}_{s, e}$.
As $b''>b$, it contradicts that $\bS^b_{s, e}\ge\bS_{s, e}-2\log T$. 
Similar arguments are applicable when $\bS^t_{s, e}$ is monotonic and therefore the lemma follows.
\hfill$\square$

\begin{lem}
\label{lem:five}
Under (\ref{lem:cond:one}) and (\ref{lem:cond:two}),
$
\mathbf{P}\left(\left|\Y^b_{s, e}\right|<\tau T^{\theta}\sqrt{\log T} \cdot n^{-1}\sumse\tystt\right)
\longrightarrow 0
$ for $b=\arg\max_{s<t<e}|\Y^t_{s, e}|$.
\end{lem}
\noindent \textbf{Proof.}
From Lemma \ref{lem:four}, there exists some $r$ such that $|b-\eta_{p_0+r}|<C\ept$.
Denote $\td=\Y^b_{s, e}=\tdo-\tdt$ and $\tm=n^{-1/2}\sumse\tystt=c_1\tdo+c_2\tdt$, where
\begin{eqnarray*}
\tdo=\sumsb\tystt, \ \
\tdt=\sumbe\tystt, \ \mbox{ and} \ c_1=c_2^{-1}=\sqrt{\frac{b-s+1}{e-b}}.
\end{eqnarray*}
For simplicity, let $c_2>c_1$. Further, let $\mui=\E\tdi$ and $\wi=\var(\tdi)$ for $i=1, 2$, and define $\mu=\E\td$ and $w=\var(\td)$.
Finally, $\thr$ denotes the threshold $\tau T^{\theta}\sqrt{\log T/n}$.
We need to show
$
\mathbf{P}(|\td|\le\tm\cdot\thr) \rightarrow 0.
$
Note that $w_i\le c^2\sup_z\sigma^4(z)\rho^2_{\infty}$.
Using Markov's and the Cauchy-Schwarz inequalities, we bound $\mathbf{P}(\td\le\tm\cdot\thr)$ by
\begin{eqnarray*}
&&\mathbf{P}\left\{(\tdo-\muo)(c_1\thr-1)+(\tdt-\mut)(c_2\thr+1)+2c_1\thr\muo+(c_2-c_1)\thr\mut\ge(1+c_1\thr)\mu\right\} \\
&&\le
4\mu^{-2}(1+c_1\thr)^{-2}\left\{(c_1\thr-1)^2\wo+(c_2\thr+1)^2\wt+4c_1^2\thr^2\muo^2+(c_2-c_1)^2\thr^2\mu_2^2\right\}
\\
&&\le 
O\left\{\mu^{-2}\sup_z\sigma^4(z)\left(\rho^2_{\infty}+\tau^2T^{2\theta}\log T\right)\right\},
\end{eqnarray*}
and since
$
\mu=\bS^b_{s, e}=O\left(\delt/\sqrt{T}\right) > T^{\theta}\sqrt{\log T},
$
the conclusion follows.
\hfill$\square$

\begin{lem}
\label{lem:six}
For some positive constants $C, \ C'$, let $s$, $e$ satisfy either
\begin{itemize}
\item[(i)] $\exists\, 1 \le p \le B$ such that $s \le \eta_p \le e$ and $[\eta_p-s+1] \wedge [e-\eta_p] \le C\ept$\, or
\item[(ii)] $\exists\, 1 \le p \le B$ such that $s \le \eta_p < \eta_{p+1} \le e$ and $[\eta_p-s+1] \vee [e-\eta_{p+1}] \le C'\ept$.
\end{itemize}
Then for a large $T$,
\[
\mathbf{P}\left(
\left|\Y^b_{s, e}\right|>\tau T^{\theta}\sqrt{\log T} \cdot n^{-1}\sumse\tystt
\right)\longrightarrow 0,
\] 
where $b=\arg\max_{s<t<e}|\Y^t_{s, e}|$.
\end{lem}
\noindent \textbf{Proof.}
First we assume (i).
Let
$
\mathcal{A}=\left\{
\left|\Y^b_{s, e}\right|> \tau T^{\theta}\sqrt{\log T} \cdot n^{-1}\sumse\tystt
\right\}
$ and
\begin{eqnarray*}
\mathcal{B}=\left\{
\frac{1}{n}\left|\sumse\left(\tystt-\E\tystt\right)\right|<
h=\frac{(\eta_p-s+1)\sigma^2_1+(e-\eta_p)\sigma^2_2}{2n}
\right\},
\end{eqnarray*}
where $\sigma^2_1=\sigma^2\left(\eta_p/T\right)$ and  $\sigma^2_2=\sigma^2\left((\eta_p+1)/T\right)$.
We have
$
\mathbf{P}\left(\mathcal{A}\right)=
\mathbf{P}\left(\mathcal{A}\cap\mathcal{B}\right)
+\mathbf{P}\left(\mathcal{A}\left|\mathcal{B}^c\right.\right)\mathbf{P}\left(\mathcal{B}^c\right)
\le\mathbf{P}\left(\mathcal{A}\cap\mathcal{B}\right)+\mathbf{P}\left(\mathcal{B}^c\right).
$
The first part is bounded as
\begin{eqnarray}
\mathbf{P}\left(\mathcal{A}\cap\mathcal{B}\right)
\le \mathbf{P}\left(\left|\Y^b_{s, e}\right|> \tau T^{\theta}\sqrt{\log T} \cdot n^{-1}\sumse\left(\E\tystt-h\right)\right).
\label{lem:six:one}
\end{eqnarray}
From Lemma \ref{lem:three}, we have $\vert \Y_{s, e}^b-\bS_{s, e}^b \vert \le \log T$.
Also Lemmas 2.2 and 2.3 of Venkatraman (1993) indicate that $\max_{s<t<e}|\bS^t_{s, e}|=|\bS^{\eta_p}|=O(\sqrt{n^{-1}\ept(n-C\ept)})=O(\sqrt{\ept})$.
Therefore $|\Y_{s, e}^b|\le |\bS^{\eta_p}|+\log T=O(\sqrt{\ept})$ and
(\ref{lem:six:one}) is bounded by
$
\E\left(\Y_{s, e}^b\right)^2/(\tau^2 h^2T^{2\theta}\log T)
\le O\left(T^{1/2-2\theta}\right) \longrightarrow 0,
$ by applying Markov's inequality.
Turning our attention to $\mathbf{P}\left(\mathcal{B}^c\right)$, we need to show that
\[
\mathbf{P}\left(\frac{1}{n}\left\vert\sum_{t=s}^e\sst(\zstt-1)\right\vert>h\right)
\longrightarrow 0.
\]
This can be shown by applying Bernstein's inequality as in the proof of Lemma \ref{lem:three}, and the lemma follows.
Similar arguments are applied when (ii) holds.
\hfill$\square$

We now prove Theorem 1. At the start of the algorithm, as $s=0$ and
$e=T-1$, all conditions for Lemma \ref{lem:five} are met and it
finds a breakpoint within the distance of $C\ept$ from the true
breakpoint, by Lemma \ref{lem:four}. Under Assumption 2,
both (\ref{lem:cond:one}) and (\ref{lem:cond:two}) are satisfied
within each segment until every breakpoint in $\sst$ is identified.
Then, either of two conditions (i) or (ii) in Lemma \ref{lem:six} is
met and therefore no further breakpoint is detected with probability
converging to 1.

Next we study how the bias present in $\E\iitt(=\sstt)$ affects
the consistency. First we define the autocorrelation wavelet
$\Psi_i(\tau)=\sumk \psi_{i, k}\psi_{i, k+\tau}$, the
autocorrelation wavelet inner product matrix $A_{i,
j}=\sum_{\tau}\Psi_i(\tau)\Psi_j(\tau)$, and the across-scales
autocorrelation wavelets $\Psi_{i, j}(\tau)=\sum_k \psi_{i,
k}\psi_{j, k+\tau}$. Then it is shown in \citet{piotr2006a} that the
integrated bias between $\E\iitt$ and $\beta_i(t/T)$ converges to
zero.

\begin{prop}[Propositions 2.1-2.2 \citep{piotr2006a}]
\label{prop:one}
Let $\iitt$ be the wavelet periodogram at a fixed scale $i$.
Under Assumption 1,
\begin{eqnarray}
T^{-1}\sum_{t=0}^{T-1}\left|\E\iitt-\beta_i(t/T)\right|^2=O(T^{-1}2^{-i})+b_{i, T}, \label{eq:prop:two}
\end{eqnarray}
where $b_{i, T}$ depends on the sequence $\{L_i\}_i$.
Further, each $\biz$ is a piecewise constant function with at most $B$ jumps, all of which occur in the set $\mathcal{B}$.
\end{prop}

Suppose the interval $[s, e]$ includes a true breakpoint $\eta_{p}$
as in (\ref{lem:cond:one}), and denote $b=\arg\max_{t\in(s, e)}
\vert\bS^t_{s, e}\vert$ and $\wh{b}=\arg\max_{t\in(s,e)}
\left\vert\bhS^t_{s, e}\right\vert$. Recall that $\E\iitt$ remains
constant within each stationary segment, apart from short (of length
$C2^{-i}$) intervals around the discontinuities in $\beta_i(t/T)$.
Suppose a jump occurs at $\eta_p$ in $\beta_i(t/T)$ yet there is no
change in $\E\iitt$ for $t\in[\eta_p-C2^{-i},\eta_p+C2^{-i}]$. Then
the integrated bias is bounded from below by $C\delt/T$ from
Assumption 2, and Proposition \ref{prop:one} is
violated. Therefore there will be a change in $\E\iitt$ as well on
such intervals around $\eta_p$ and $\E I^{(i)}_{t_1, T}\ne\E
I^{(i)}_{t_2, T}$ for $t_1\le\eta_p-C2^{-i}$ and
$t_2\ge\eta_p+C2^{-i}$. Although the bias of $\E\iitt$ in relation
to $\beta_i(t/T)$ may cause some bias between $\wh{b}$ and $b$, we
have that $|\wh{b}-b|\le C2^{\is}<\ept$ holds for $\is=O(\log\log
T)$, which is an admissible rate for $\is$. Besides, once one
breakpoint is detected in such intervals, the algorithm does not
allow any more breakpoints to be detected within the distance of
$\Delt$ from the detected breakpoint, by construction. Hence the
bias in $\E\iitt$ does not affect the results of Lemmas
\ref{lem:one}--\ref{lem:six} for wavelet periodograms at finer
scales and the consistency still holds for $\ystt$ in (3).

Finally, we note that the within-scale post-processing step in Section 3.2.1 is in line with the theoretical consistency of our procedure; (a) Lemma \ref{lem:five} implies that our test statistic exceeds the threshold when there is a breakpoint $\eta$ within a segment $[s, e]$ which is of sufficient distance from both $s$ and $e$, and (b) Lemma \ref{lem:six} shows that it does not exceed the threshold when $(s, \eta, e)$ does not satisfy the condition in (a).

\section{The proof of Theorem 2}
\label{sec:proof:two}

From Assumption 1 and the invertibility of the
autocorrelation wavelet inner product matrix $A$, there exists at
least one sequence of wavelet periodograms among $\iitt, \ i=-1,
\ldots, -\is$ in which any breakpoint in $\mathcal{B}$ is detected.
Suppose there is only one such scale, $i_0$, for
$\nu_q\in\mathcal{B}$ and denote the detected breakpoint as
$\heta^{(i_0)}_{p_0}$. After the across-scales post-processing,
$\heta^{(i_0)}_{p_0}$ is selected as $\wh{\nu}_q$ since no other
$\heta^{(i)}_p$, $i\ne i_0$, is within the distance of $\Lamt=C\ept$
from either $\wh{\nu}_q$ or $\heta^{(i_0)}_{p_0}$, and
$\left\vert\nu_q-\heta^{(i_0)}_{p_0}\right\vert\le\ept$ with
probability converging to 1 from Theorem 1. If there are
$D(\le\is)$ breakpoints detected for $\nu_q$, denote them as
$\heta^{(i_1)}_{p_1}, \ldots, \heta^{(i_D)}_{p_D}$. Then for any
$1\le a<b \le D$, $
\left\vert\heta^{(i_a)}_{p_a}-\heta^{(i_b)}_{p_b}\right\vert \le
\left\vert\heta^{(i_a)}_{p_a}-\nu_q\right\vert +
\left\vert\heta^{(i_b)}_{p_b}-\nu_q\right\vert \le C\ept, $ and only
the one from the finest scale is selected as $\wh{\nu}_q$ among
them by the post-processing procedure. Hence the across-scales
post-processing preserves the consistency for the breakpoints
selected as its outcome.

\end{document}